\documentclass[%
 aip,
 jmp,%
 amsmath,amssymb,
reprint,%
]{revtex4-1}

\usepackage[colorlinks,linkcolor=blue,citecolor=blue,urlcolor=black,bookmarks=false,hypertexnames=true]{hyperref} 

\usepackage{amsmath}
\usepackage{xspace}
\usepackage{multirow}
\usepackage{latexsym}
\usepackage{braket}
\usepackage{cleveref}
\usepackage{threeparttable}
\usepackage{enumerate}
\usepackage{mathrsfs}
\usepackage{enumitem}
\usepackage{color}
\usepackage[version=3]{mhchem}
\usepackage{caption,setspace}
\usepackage{graphicx}
\usepackage{subfigure}
\usepackage{threeparttable}

\allowdisplaybreaks
\usepackage{array}
\newcolumntype{L}[1]{>{\raggedright\let\newline\\\arraybackslash\hspace{0pt}}m{#1}}
\newcolumntype{C}[1]{>{\centering\let\newline\\\arraybackslash\hspace{0pt}}m{#1}}
\newcolumntype{R}[1]{>{\raggedleft\let\newline\\\arraybackslash\hspace{0pt}}m{#1}}

\crefname{figure}{Figure}{Figures}
\crefname{table}{Table}{Tables}
\crefname{equation}{Eq.}{Eqs.}
\crefname{section}{Section}{Sections}

\begin{document}

\author{Samragni~Banerjee}
\affiliation{Department of Chemistry and Biochemistry, The Ohio State University, Columbus, Ohio 43210, USA}

\author{Alexander~Yu.~Sokolov}
\email{sokolov.8@osu.edu}
\affiliation{Department of Chemistry and Biochemistry, The Ohio State University, Columbus, Ohio 43210, USA}

\title{
\color{blue}
Efficient implementation of the single-reference algebraic diagrammatic construction theory for charged excitations: Applications to the TEMPO radical and DNA base pairs
\vspace{0.25cm}
}

\begin{abstract}
We present an efficient implementation of the second- and third-order single-reference algebraic diagrammatic construction theory for electron attachment (EA) and ionization (IP) energies and spectra (EA/IP-ADC($n$), $n = 2, 3$). Our new EA/IP-ADC program features spin adaptation for closed-shell systems, density fitting for efficient handling of the two-electron integral tensors, as well as vectorized and parallel implementation of tensor contractions. We demonstrate capabilities of our efficient implementation by applying the EA/IP-ADC($n$) (n = 2, 3) methods to compute the photoelectron spectrum of the TEMPO radical, as well as the vertical and adiabatic electron affinities of TEMPO and two DNA base pairs (guanine--cytosine and adenine--thymine). The spectra and electron affinities computed using large diffuse basis sets with up to 1028 molecular orbitals are found to be in a good agreement with the best available results from the experiment and theoretical simulations.
\end{abstract}

\titlepage

\maketitle

\section{Introduction}
\label{sec:intro}
Accurate theoretical simulations of charged electronic states are crucial for reliable predictions of many important properties of molecules and materials, such as electron affinities (EA), ionization potentials (IP), band gaps, and photoelectron spectra.\cite{Cederbaum:1975p290,VonNiessen:1984p57,Ortiz:2012p123,Hirata:2015p1595} Computations of charged excitations face many challenges associated with the description of charge distribution and open-shell electronic states that require accurate treatment of orbital relaxation and electron correlation effects. Reliable simulations of electron attachment and ionization can be performed using high-level ab initio methods, such as state-specific or equation-of-motion (EOM) coupled cluster theory (CC)\cite{Nooijen:1992p55,Nooijen:1993p15,Nooijen:1995p1681,Stanton:1993p7029,Krylov:2008,Kowalski:2014p094102,BhaskaranNair:2016p144101,Peng:2018p4335} and multireference configuration interaction,\cite{Fleig:2008p270,Streit:2011p22} but these methods are usually limited to small systems due to their high computational cost. On the other hand, computationally efficient approaches, such as those based on time-dependent many-body perturbation theory, often combined with density functional theory,\cite{Mckechnie:2015p194114,Nguyen:2012p081101,Hedin:1965p796,Faleev:2004p126406,vanSchilfgaarde:2006p226402,Neaton:2006p216405,Samsonidze:2011p186404,vanSetten:2013p232,Reining:2017pe1344} can be applied to larger systems but may be unreliable and are difficult to improve systematically.
The development of accurate yet efficient computational methods for charged electronic states is an ongoing challenge and is an active area of research.\cite{Pavosevic:2017p121101,Lange:2018p4224,Backhouse:2020p6294,Dempwolff:2019p064108,Dempwolff:2020p024113,Dempwolff:2020p024125}

An attractive alternative to traditional coupled cluster theory for simulating charged electronic states is algebraic diagrammatic construction (ADC) theory\cite{Schirmer:1982p2395,Schirmer:1983p1237,Schirmer:1991p4647,Mertins:1996p2140,Schirmer:2004p11449,Dreuw:2014p82} that offers an efficient approach for calculating excitation energies and transition intensities from poles and residues of a many-body propagator approximated in a time-independent perturbation series. Since its formulation in 1982, single-reference ADC theory has been applied to a wide range of problems in chemistry and spectroscopy, including simulating UV/Vis\cite{Schirmer:1982p2395,Starcke:2009p024104,Harbach:2014p064113} and X-ray absorption,\cite{Barth:1985p867,Wenzel:2014p1900} photoelectron,\cite{Schirmer:1983p1237,Schirmer:1998p4734,Trofimov:2005p144115,Angonoa:1987p6789,Schirmer:2001p10621,Thiel:2003p2088,Dempwolff:2019p064108,Dempwolff:2020p024113,Dempwolff:2020p024125} and two-photon absorption spectra.\cite{Knippenberg:2012p064107} The ADC methods for charged excitations (EA/IP), first proposed in 1983, were based on the Dyson framework that involves perturbative self-energy expansion of the Dyson equation and couples the EA and IP components of the one-electron propagator.\cite{Schirmer:1983p1237} Later work by Schirmer, Trofimov, and co-workers\cite{Schirmer:1998p4734,Trofimov:2005p144115} established the non-Dyson EA/IP-ADC formalism, where the EA and IP components of the propagator are decoupled. Recently, our group developed a multireference formulation of the non-Dyson EA/IP-ADC theory for simulations of charged excitations in strongly correlated chemical systems.\cite{Sokolov:2018p204113,Chatterjee:2019p5908,Chatterjee:2020p6343} 

Several implementations of the single-reference non-Dyson EA/IP-ADC methods were reported in the literature, most of them limited to calculations of only ionization processes (IP). In 2005, Trofimov and Schirmer presented the first pilot implementation of the non-Dyson IP-ADC($n$) method up to third order in perturbation theory. More recently, two efficient implementations of non-Dyson IP-ADC($n$) ($n \le 3$) were reported by Schneider et al.\cite{Schneider:2015p144103} and Dempwolff et al.\cite{Dempwolff:2019p064108,Dempwolff:2020p024113,Dempwolff:2020p024125} and the latter is available in the \textsc{Q-Chem} package.\cite{qchem:44} 
Implementations of the non-Dyson EA-ADC($n$) ($n \le 3$) methods were limited to two early benchmark studies,\cite{Trofimov:2011p77,Schneider:2015} a more extensive benchmark of EA-ADC($n$) ($n \le 3$) for closed- and open-shell systems performed in our group,\cite{Banerjee:2019p224112} and a recent efficient implementation of EA-ADC(2) by Liu et al.\cite{Liu:2020p174109} To the best of our knowledge, an efficient implementation of EA-ADC(3) has not been reported. While applications of non-Dyson IP-ADC(3) were presented for systems with up to 639 molecular orbitals,\cite{Dempwolff:2019p064108,Dempwolff:2020p024113} the largest non-Dyson EA-ADC(3) calculations reported to date was limited to 298 orbitals.\cite{Banerjee:2019p224112}
 
In this work, we present an efficient implementation of the non-Dyson EA- and IP-ADC($n$) methods ($n\leq3$), now available as a module in the \textsc{PySCF} software package.\cite{Sun:2020p024109} Our EA/IP-ADC($n$) program features the spin-restricted (RADC) and unrestricted (UADC) codes for closed- and open-shell systems, respectively, takes advantage of the vectorized and parallel implementation of tensor contractions, and employs the density fitting approximation\cite{Whitten:1973p4496,Vahtras:1993p514,Feyereisen:1993p359,Rendell:1994p400,Weigend:2002p4285} for efficient disk and memory management. We demonstrate the efficiency of our implementation by performing the EA- and IP-UADC(3) calculations for the open-shell (2,2,6,6-tetramethylpiperidin-1-yl)oxyl (TEMPO) radical with up to 758 orbitals and EA-RADC(3) calculations for the closed-shell DNA base pairs with up to 1028 basis functions, which are the largest non-Dyson EA/IP-ADC(3) calculations reported to date.

We begin our discussion with a brief overview of ADC (\cref{sec:theory}), followed by the overview of our implementation (\cref{sec:implementation}), where we discuss the general algorithm (\cref{sec:implementation:overview}) along with the details of spin adaptation (\cref{sec:implementation:spin-adaptation}) and the density fitting approximation (\cref{sec:implementation:density-fitting}). We next present computational details in \cref{sec:comp_details} and benchmark the accuracy of the density fitting approximation along with the efficiency of our implementation in \cref{sec:results:accuracy_df}. Finally, we present our results for the TEMPO radical in \cref{sec:results:TEMPO} and two DNA base pairs (guanine--cytosine, adenine--thymine) in \cref{sec:results:base_pairs}. We present our conclusions in \cref{sec:conclusions}.
 
\section{Theory}
\label{sec:theory}

\subsection{Non-Dyson ADC theory for the one-particle propagator}
\label{sec:theory:general_overview}

We first give a brief overview of the non-Dyson algebraic diagrammatic construction (ADC) theory\cite{Schirmer:1982p2395,Schirmer:1983p1237,Schirmer:1991p4647,Mertins:1996p2140,Schirmer:2004p11449,Dreuw:2014p82,Banerjee:2019p224112} for electron attachment and removal, where the central object of interest is the retarded single-particle Green's function
\begin{align}
G_{pq}(\omega) &= G_{pq}^{+}(\omega) + G_{pq}^{-}(\omega)\notag\\
               &=\langle\Psi_{0}^{N}|a_{p} (\omega - H + E_{0}^{N})^{-1}a_{q}^{\dagger}|\Psi_{0}^{N}\rangle\notag\\
               &+ \langle\Psi_{0}^{N}|a_{q}^{\dagger} (\omega + H - E_{0}^{N})^{-1}a_{p}|\Psi_{0}^{N}\rangle\label{eq: G_pq}
\end{align}
Here, the forward ($G_{pq}^{+}(\omega)$)  and backward ($G_{pq}^{-}(\omega)$) components of the propagator describe electron attachment (EA) and ionization (IP) processes, respectively, $|\Psi_{0}^{N}\big\rangle$ and  \textit{E}$_{0}^{N}$ are the eigenfunction and eigenvalue of the electronic Hamiltonian  \textit{H} of $N$-electron system, and a frequency \mbox{$\omega \equiv \omega' + i\eta$} is expressed in terms of its real ($\omega'$) and infinitesimal imaginary ($i\eta$) components.  The operators $a_{p}^{\dag}$ and $a_{p}$ are the usual fermionic creation and annihilation operators, respectively. 

The forward and backward propagators in \cref{eq: G_pq} can be expressed in the matrix form
\begin{align}
    \textbf{G}_{\pm}(\omega) = \mathbf{\tilde{X}}_{\pm}(\omega\textbf{1} - \boldsymbol{\tilde{\Omega}}_{\pm})^{-1}\mathbf{\tilde{X}}_{\pm}^{\dagger}\label{eq: G_pq_matrix}
\end{align}
where $\boldsymbol{\tilde{\Omega}}_{\pm}$ are the diagonal matrices of the exact vertical attachment ($\omega_{n} = E_{n}^{N+1} - E_{0}^{N}$) and ionization ($\omega_{n} = E_{0}^{N} - E_{n}^{N - 1}$) energies and $\mathbf{\tilde{X}}_{\pm}$ are the matrices of the spectroscopic amplitudes with elements $\tilde{X}_{+pn} = \langle\Psi_{0}^{N}|a_{p}|\Psi_{n}^{N+1}\rangle$ and $\tilde{X}_{-qn} = \langle\Psi_{0}^{N}|a_{q}^{\dagger}|\Psi_{n}^{N-1}\rangle$, where $\ket{\Psi_{n}^{N+1}}$ and $\ket{\Psi_{n}^{N-1}}$ are the exact eigenstates and $E_{n}^{N+1}$ and $E_{n}^{N-1}$ are the exact eigenvalues of the Hamiltonian $H$ for the $(N+1)$- and $(N-1)$-electron system, respectively. \cref{eq: G_pq_matrix} is known as the spectral (or Lehmann) representation of the propagator.\cite{Kobe:1962p448}

The propagator in \cref{eq: G_pq,eq: G_pq_matrix} expressed in the basis of the exact eigenstates $\ket{\Psi_{n}^{N+1}}$ and $\ket{\Psi_{n}^{N-1}}$ is computationally expensive. In non-Dyson ADC, the propagator is rewritten in an approximate (in general, non-orthogonal) basis of the ($N\pm 1$)-electron states
\begin{align}
    \textbf{G}_{\pm}(\omega) = \textbf{T}_{\pm}(\omega\textbf{S}_{\pm}-\textbf{M}_{\pm})^{-1}\textbf{T}_{\pm}^{\dagger}\label{eq: G_approx}
\end{align}
where the $\textbf{M}_{\pm}$ matrices are no longer diagonal, in contrast to $\boldsymbol{\tilde{\Omega}}_{\pm}$ in \cref{eq: G_pq_matrix}. In \cref{eq: G_approx}, $\textbf{M}_{\pm}$ and $\textbf{T}_{\pm}$, known as the effective Hamiltonian and transition moment matrices, contain information about charged excitation energies and probabilities of electronic transitions, respectively. Since the approximate basis states are in general non-orthogonal, information about their overlap is contained in the $\textbf{S}_{\pm}$ matrices. For all single-reference ADC methods discussed in this work, the approximate basis is orthonormal and, thus, we will assume $\textbf{S}_{\pm} = \textbf{1}$ henceforth.

Starting with \cref{eq: G_approx}, the approximate propagator $\textbf{G}_{\pm}(\omega)$ is expanded in a perturbative series 
 \begin{align} 
   \textbf{G$_{\pm}(\omega)$} &= \textbf{G$_{\pm}^{(0)}(\omega)$} +\textbf{G$_{\pm}^{(1)}(\omega)$} + ... + \textbf{G$_{\pm}^{(n)}(\omega)$} + ... \label{eq: G_n}
\end{align}
Truncating this expansion at the $n$-th order defines the propagator of the $n$-th order non-Dyson ADC approximation for EA or IP (EA/IP-ADC($n$)). Working equations for $ \textbf{G$_{\pm}(\omega)$} $ are derived by evaluating $\textbf{M}_{\pm}$ and $\textbf{T}_{\pm}$ up to the $n$-th order in perturbation theory using the intermediate state representation approach\cite{Schirmer:1991p4647,Mertins:1996p2140,Schirmer:2004p11449} or via the formalism of the effective Liouvillean theory.\cite{Mukherjee:1989p257,Sokolov:2018p204113,Banerjee:2019p224112} 

In the following, we briefly overview the derivation of the non-Dyson EA/IP-ADC($n$) equations using effective Liouvillean theory. In this approach, the exact ground-state wavefunction is expressed via a unitary transformation
 \begin{align}
|\Psi_{0}^{N}\rangle = e^{T - T^{\dagger}}|\Phi\rangle\label{eq: gs_wavefunction}
\end{align}
 where $|\Phi\rangle$ is a reference Slater determinant and the operator $T$ generates all possible excitations from the occupied to virtual (external) orbitals labeled using the $i,j,k,l,\ldots$ and $a,b,c,d,\ldots$ indices, respectively 
 \begin{align}
    T = \sum_m^N T_{m}, \quad T_{m} = \frac{1}{(m!)^{2}}\sum_{ijab\ldots} t_{ij\ldots}^{ab\ldots}a_{a}^{\dagger}a_{b}^{\dagger}\ldots a_{j}a_{i}\label{eq: ex_op}
\end{align}
To define perturbative series \eqref{eq: G_n}, the electronic Hamiltonian $H$ is partitioned into the zeroth-order part $H^{(0)}$ and a perturbation $V = H -H^{(0)}$. This allows to determine contributions to $\textbf{M}_{\pm}$ and $\textbf{T}_{\pm}$ at each order in perturbation theory. 

The $n$th-order contributions to the EA-ADC matrices have the form:
\begin{align}
    M_{+\mu\nu}^{(n)} &= \sum_{klm}^{k+l+m=n} \langle \Phi|[h_{+\mu}^{(k)},[\tilde{H}^{(l)},h_{+\nu}^{(m)\dagger}]]_{+}|\Phi\rangle \label{eq: M_ea} \\
    T_{+p\nu}^{(n)} &= \sum_{kl}^{k+l=n} \langle \Phi|[\tilde{a}_{p}^{(k)},h_{+\nu}^{(l)\dagger}]_{+}|\Phi\rangle \label{eq: T_ea}
\end{align}
where operators $\tilde{H}^{(k)}$ and $\tilde{a}_{p}^{(k)}$ are the $k$-th-order contributions to the effective Hamiltonian $\tilde{{H}}=e^{-(T-T^{\dagger})}He^{(T-T^{\dagger})}$ and observable $\tilde{a}_{p}=e^{-(T-T^{\dagger})}a_{p}e^{(T-T^{\dagger})}$ operators, while $[\ldots]$ and $[\ldots]_{+}$ denote commutator and anticommutator, respectively. 
The operators $h_{+\mu}^{(k)\dagger}$ in \cref{eq: M_ea,eq: T_ea} are used to construct a set of basis states $\ket{\Psi^{(k)}_{+\mu}} = h_{+\mu}^{(k)\dagger}\ket{\Phi}$ necessary for expanding the eigenstates of the $(N+1)$-electron system in the EA-ADC equations. Similarly, the $n$th-order contributions to the IP-ADC matrices are given as:
\begin{align}
    M_{-\mu\nu}^{(n)} &= \sum_{klm}^{k+l+m=n} \langle \Phi|[h_{-\mu}^{(k)\dagger},[\tilde{H}^{(l)},h_{-\nu}^{(m)}]]_{+}|\Phi\rangle\label{eq: M_ip}  \\
    T_{-p\nu}^{(n)} &= \sum_{kl}^{k+l=n} \langle \Phi|[\tilde{a}_{p}^{(k)},h_{-\nu}^{(l)}]_{+}|\Phi\rangle\label{eq: T_ip} 
\end{align}
where the ionization operators $h_{-\mu}^{(k)\dagger}$ define the  $(N-1)$-electron basis states $\ket{\Psi^{(k)}_{-\mu}} = h_{-\mu}^{(k)\dagger}\ket{\Phi}$. 
For the low-order EA/IP-ADC($n$) approximations ($n$ $\le$ 3), only the zeroth- and first-order components of $h_{\pm \mu}^{(k)\dagger}$ are required and are given as follows: $h_{+\mu}^{(0)\dagger} = a_{a}^{\dagger}$, ${h}_{+\mu}^{(1)\dagger} =a_b^{\dagger}a_a^{\dagger}a_{i}$, $h_{-\mu}^{(0)\dagger} = a_{i}$, and ${h}_{-\mu}^{(1)\dagger} =a_a^{\dagger}a_ja_{i}$. The $\tilde{H}^{(k)}$ and $\tilde{a}_{p}^{(k)}$ operators in \cref{eq: M_ea,eq: T_ea,eq: M_ip,eq: T_ip} are obtained by expanding $\tilde{H}$ and $\tilde{a}_{p}$ using the Baker--Campbell--Hausdorff (BCH) formula
\begin{align}
    \tilde{H} &= H^{(0)} + V + [H^{(0)},T^{(1)} - T^{(1){\dagger}}] + [H^{(0)},T^{(2)} - T^{(2){\dagger}}]\notag   \\
    &+\frac{1}{2!}[V+(V + [H^{(0)},T^{(1)} - T^{(1){\dagger}}]),T^{(1)}-T^{(1){\dagger}}]+...\label{eq: H_bch}\\ 
   \tilde{a}_p &= a_p + [a_p,T^{(1)}-T^{(1){\dagger}}]+ [a_p,T^{(2)}-T^{(2){\dagger}}]\notag\\
    &+\frac{1}{2!}[[a_p,T^{(1)}-T^{(1){\dagger}}],T^{(1)}-T^{(1){\dagger}}]+...\label{eq: ap_bch}
\end{align}
and collecting terms at the $k$-th order. These equations depend on the amplitudes of the excitation operators $T^{(k)}$ (\cref{eq: ex_op}). The low-order EA/IP-ADC($n$) ($n$ $\le$ 3) approximations require calculating up to the $n$-th order single-excitation amplitudes ($t_{i}^{a(k)}$) and up to the ($n-1$)-th order double-excitation amplitudes ($t_{ij}^{ab(k-1)}$) . At each order $k$, these amplitudes are computed by solving a system of projected amplitude equations:
\begin{align}
    \langle\Phi|a_{i}^{\dagger}a_{a}\tilde{H}^{(k)}|\Phi\rangle = 0\label{eq: t_singles} \\
    \langle\Phi|a_{i}^{\dagger}a_{j}^{\dagger}a_{b}a_{a}\tilde{H}^{(k)}|\Phi\rangle = 0 \label{eq: t_doubles}
\end{align}
The resulting amplitudes are equivalent to those that define the $k$-th-order wavefunction in the single-reference M\o ller--Plesset perturbation theory.

Once the equations for $\textbf{M}_{\pm}$ and $\textbf{T}_{\pm}$ are obtained, the EA/IP-ADC($n$) transition energies are computed by solving the eigenvalue problem
\begin{align}
    \textbf{M}_{\pm}\textbf{Y}_{\pm}=\textbf{Y}_{\pm}\boldsymbol{\Omega}_{\pm}\label{eq: eig}
\end{align}
where $\boldsymbol\Omega_{\pm}$ is a diagonal matrix of eigenvalues and \textbf{Y}$_{\pm}$ is a matrix of eigenvectors that are used to compute the spectroscopic amplitudes 
\begin{align}
    \textbf{X}_{\pm}=\textbf{T}_{\pm} \textbf{Y}_{\pm}\label{eq: spec_amp}
\end{align}
which provide information about transition intensities. The elements of $\textbf{X}_{\pm}$ can be used to calculate the spectroscopic factors
\begin{align}
	\label{eq:spec_factors}
	P_{\pm\mu} = \sum_{p} |X_{\pm p\mu}|^2
\end{align}
as well as the ADC propagator and density of states
\begin{align}
	\label{eq: g_adc}
	\mathbf{G}_{\pm}(\omega) &= \mathbf{X}_{\pm} \left(\omega \mathbf{1} - \boldsymbol{\Omega}_{\pm}\right)^{-1}  \mathbf{X}_{\pm}^\dag \\
	\label{eq: spec_function}
	A(\omega) &= -\frac{1}{\pi} \mathrm{Im} \left[ \mathrm{Tr} \, \mathbf{G}_{\pm}(\omega) \right]
\end{align}
Working equations for the non-Dyson EA/IP-ADC($n$) ($n$ = 2, 3) methods in the spin-orbital basis have been presented in Ref.\@ \citenum{Banerjee:2019p224112}.

\section{Implementation}
\label{sec:implementation}

\subsection{Overview of the EA/IP-ADC implementation}
\label{sec:implementation:overview}

We have implemented the EA/IP-ADC($n$) (n = 2, 3) methods as a module in the \textsc{PySCF} program.\cite{Sun:2020p024109} Our efficient implementation consists of the spin-restricted (EA/IP-RADC) and unrestricted (EA/IP-UADC) codes, for closed- and open-shell systems, respectively. The RADC($n$)/UADC($n$) computation is preceded by the corresponding restricted/unrestricted (RHF/UHF) Hartree--Fock calculation and follows the algorithm described below:

\begin{table*}[t!]
	\captionsetup{justification=raggedright,singlelinecheck=false,font=footnotesize}
	\caption{Comparison of formal computational scaling of the EA/IP-ADC($n$) methods ($n \le 3$) for each step of the algorithm outlined in \cref{sec:implementation:overview}. $O$ and $V$ denote the number of occupied and virtual molecular orbitals, respectively. Also shown are the corresponding computational prefactors relative to those of RADC(3).}
	\label{tab: scaling}
	\footnotesize
	\setstretch{1}
    \begin{tabular}{L{4.5cm}C{2cm}C{1.5cm}C{2cm}C{1.5cm}C{1.5cm}C{1.5cm}}
       \hline
         \hline
        \multicolumn{1}{c}{Component} &\multicolumn{2}{c}{EA}   &\multicolumn{2}{c}{IP}  &\multicolumn{2}{c}{Relative prefactor} \\
        &ADC(2) &ADC(3) &ADC(2) &ADC(3) &RADC(3) &UADC(3)\\ 
        \hline
Amplitudes and MP$n$ energy									&$O^{2}V^{3}$		&$O^{2}V^{4}$		&$O^{2}V^{3}$		&$O^{2}V^{4}$		&1 		&3\\   	               
\textbf{M}$_{ab}$(1p-1p) or \textbf{M}$_{ij}$(1h-1h)     					&$O^{2}V^{3}$		&$O^{2}V^{4}$		&$O^{3}V^{2}$		&$O^{2}V^{4}$		&1 		&3 \\   
$\boldsymbol{\sigma}_{\pm}^{(i+1)} = \mathbf{M_{\pm}Y_{\pm}}^{(i)}$	&$OV^{3}$		&$O^{2}V^{3}$		&$O^{3}V$		&$O^{3}V^{2}$		&1		&4 \\
Spectroscopic factors  ($P_{\pm\mu}$)							&$O^{2}V^{3}$		&$O^{2}V^{3}$		&$O^{3}V^{2}$		&$O^{3}V^{2}$		&1		&3 \\
\hline
Overall													&$O^{2}V^{3}$		&$O^{2}V^{4}$		&$O^{2}V^{3}$		&$O^{2}V^{4}$		&1		&3 \\
       \hline
        \hline
    \end{tabular}
\end{table*}

\begin{enumerate}
\item Transform two-electron repulsion integrals $(pq|rs)$ from the atomic to the RHF/UHF molecular orbital basis.
\item Compute the amplitudes of the effective Hamiltonian ($t_{i}^{a(k)}$ and $t_{ij}^{ab(k-1)}$, $k \le n$) by solving \cref{eq: t_singles,eq: t_doubles} and evaluate the $n$-th-order M\o ller-Plesset (MP$n$) correlation energy, where $n$ is the order of the ADC($n$) approximation.
\item Compute small blocks of the effective Hamiltonian matrix $\textbf{M}_{\pm}$ with respect to the zeroth-order operators ($h_{\pm\mu}^{(0)\dagger}$, \cref{sec:theory:general_overview}). These are the 1p-1p ($M_{+ab}$, \cref{eq: M_ea}) and 1h-1h ($M_{-ij}$, \cref{eq: M_ip}) blocks of $\textbf{M}_{+}$ and $\textbf{M}_{-}$ for EA- and IP-ADC($n$), respectively.
\item Define a function for calculating a matrix-vector product $\boldsymbol{\sigma}_\pm = \mathbf{M}_\pm \mathbf{Y}_\pm$ using the approximate eigenvectors $\mathbf{Y}_\pm$ and the small blocks of the effective Hamiltonian matrix precomputed in step 3. Solve the eigenvalue problem \eqref{eq: eig} by optimizing the eigenvectors $\mathbf{Y}_\pm$ until convergence using an iterative diagonalization algorithm to obtain low-energy EA/IP's. 
\item Using the converged eigenvectors $\mathbf{Y}_\pm$, compute spectroscopic amplitudes $\mathbf{X}_\pm$ (\cref{eq: spec_amp}) and spectroscopic factors $P_{\pm\mu}$ (\cref{eq:spec_factors}). If desired, compute spectral density $A(\omega)$ (\cref{eq: g_adc}), Dyson orbitals, and other molecular properties. 
\end{enumerate}

Our efficient EA/IP-ADC($n$) program written entirely in Python takes advantage of the highly optimized Basic Linear Algebra Subroutines (BLAS) and Open Multi-Processing (OpenMP) parallelization for the most expensive tensor contractions at each step of the algorithm described above. Implementation of the inexpensive but tedious tensor contractions was performed using the capabilities of the NumPy package.\cite{Harris:2020p357} Our code features three algorithms for the efficient management of the memory and disk space: (i) {\it in-core} algorithm where all tensors such as two-electron integrals and amplitudes are stored in memory, (ii) {\it out-of-core} algorithm that stores all large tensors on disk, and (iii) {\it density-fitted} algorithm that significantly reduces disk usage and enables memory savings by approximating two-electron integrals as discussed in \cref{sec:implementation:density-fitting}. We use the multiroot Davidson algorithm\cite{Davidson:1975p87,Liu:1978p49} available in \textsc{PySCF} for iterative solution of the ADC eigenvalue problem. 

\cref{tab: scaling} shows the computational scaling of each step of the ADC algorithm with the number of occupied ($O$) and virtual ($V)$ molecular orbitals. For $n$ = 2 and 3, the overall computational scaling of EA/IP-ADC($n$) is $\mathcal{O}(O^2V^3)$ and $\mathcal{O}(O^2V^4)$, respectively. Importantly, precomputing the 1p-1p and 1h-1h blocks of the effective Hamiltonian matrix $\mathbf{M}_\pm$ at step 3 and reusing them in the Davidson diagonalization at step 4 lowers the cost of computing the  $\boldsymbol{\sigma}_\pm = \mathbf{M}_\pm \mathbf{Y}_\pm$ matrix-vector products from $\mathcal{O}(O^2V^4)$ to $\mathcal{O}(O^2V^3)$ and $\mathcal{O}(O^3V^2)$ for EA- and IP-ADC(3), respectively. As a result, EA/IP-RADC(3) and EA/IP-UADC(3) require a total of two and six $\mathcal{O}(O^2V^4)$ contractions, respectively, independent of the number of iterations in the Davidson algorithm. A similar approach that allows to reduce the scaling of the matrix-vector products has been employed by Dempwolff and co-workers in the efficient implementation of IP-ADC(3).\cite{Dempwolff:2019p064108} 

We note that the computational scaling of solving the EA/IP-ADC($n$) eigenvalue problem can in principle be lowered if, instead of precomputing the 1p-1p and 1h-1h blocks of the effective Hamiltonian matrix $\mathbf{M}_\pm$ at step 3, the contributions of these blocks to the matrix-vector products (i.e., $\sum_b M_{+ab}Y_b$ and $\sum_j M_{-ij}Y_j$) are recomputed at every iteration of step 4 by forming suitable efficient intermediates. This avoids step 3 and changes the scaling of step 4 to $\mathcal{O}(O^2V^2)$ for EA/IP-ADC(2) and to $\mathcal{O}(O^2V^3)$/$\mathcal{O}(OV^4)$ for EA/IP-ADC(3), respectively, but does not change the overall formal scaling of the EA/IP-ADC($n$) methods due to the higher computational scaling of step 2.

In the following, we discuss the derivation of the spin-adapted equations used in our EA/IP-RADC(3) implementation (\cref{sec:implementation:spin-adaptation}) and give an overview of the density fitting approximation used for the two-electron integrals (\cref{sec:implementation:density-fitting}). The efficiency of our implementation and the accuracy of the density fitting approximation are analyzed in \cref{sec:results:accuracy_df}.

\subsection{Spin-restricted EA/IP-ADC for closed-shell systems (EA/IP-RADC)}
\label{sec:implementation:spin-adaptation}

In our earlier work,\cite{Banerjee:2019p224112} we presented a general spin-orbital derivation and implementation of EA/IP-ADC($n$) ($n\le3$) applicable to closed- and open-shell systems. Here, we discuss spin adaptation of the EA/IP-ADC equations used in our efficient spin-restricted EA/IP-RADC implementation. While spin adaptation of the EA-ADC equations has not been discussed previously in the literature, it has been presented for other ADC formulations, such as ADC for ionization\cite{Thiel:2003p2088} and neutral\cite{Barth:1985p867,Trofimov:2002p6402} excitation energies. 

In our derivation of the EA/IP-RADC equations we follow the approach described in Ref.\@ \citenum{Nooijen:1995p3629}. 
Starting with spin-orbital EA/IP-ADC equations, we obtain the spin-adapted equations for the effective Hamiltonian ($\mathbf{M}_\pm$) and effective transition moments ($\mathbf{T}_\pm$) matrix elements in \cref{eq: M_ea,eq: T_ea,eq: M_ip,eq: T_ip} by integrating over the spin variables. We denote each spin-orbital with an index $p_\sigma$, where $\sigma = \{ \alpha, \beta \}$ is a spin label and $p$ is an index of the spatial molecular orbital (MO) from the restricted Hartree--Fock (RHF) calculation. We use the $i,j,k,l,\ldots$ and $a,b,c,d,\ldots$ indices to label the occupied and virtual MO's, respectively, and reserve $p,q,r,s,\ldots$ to denote a general (occupied or virtual) MO. 

Integrating over the spin variables allows to express the spin-orbital $\mathbf{M}_\pm$ and $\mathbf{T}_\pm$ matrix elements in terms of their counterparts that depend solely on spatial orbitals. As an example, the matrix elements of $\mathbf{M}_-$ and $\mathbf{T}_-$ in IP-ADC can be expressed as:
\begin{align}
 M_{-i_{\alpha}, j_{\alpha}} &= M_{-i_{\beta}, j_{\beta}} = M_{-i,j}  \label{eq:M_ij}\\
 M_{-i_{\alpha}, j_{\beta} k_{\alpha} a_{\beta}} &= M_{-i_{\beta}, j_{\alpha} k_{\beta} a_{\alpha}}  =  M_{-i, jka} \label{eq:M_ijka1} \\
 M_{-i_{\alpha}, j_{\alpha} k_{\alpha} a_{\alpha}} &= M_{-i_{\beta}, j_{\beta} k_{\beta} a_{\beta}}  \notag \\
 &=  M_{-i, jka} - M_{-i,kja} \label{eq:M_ijka2}\\
 M_{-i_{\beta}l_{\alpha}b_{\beta}, j_{\beta} k_{\alpha} a_{\beta}} &= M_{-i_{\alpha}l_{\beta}b_{\alpha}, j_{\alpha} k_{\beta} a_{\alpha}}  =  M_{-ilb, jka}\label{eq:M_ilbjka1}\\
 M_{-i_{\alpha}l_{\alpha}b_{\alpha}, j_{\alpha} k_{\alpha} a_{\alpha}} &= M_{-i_{\beta}l_{\beta}b_{\beta}, j_{\beta} k_{\beta} a_{\beta}} \notag  \\ 
 &=  M_{-ilb, jka} - M_{-lib, jka} \notag \\
 &- M_{-ilb, kja} + M_{-lib, kja}\label{eq:M_ilbjka2} \\
 T_{-p_\alpha,i_\alpha} &= T_{-p_\beta,i_\beta} = T_{-p,i} \label{eq:T_i}\\
T_{-p_\alpha,j_{\beta} k_{\alpha} a_{\beta}} &= T_{-p_\beta,j_{\alpha} k_{\beta} a_{\alpha}} = T_{-p,jka} \label{eq:T_jka1}\\
T_{-p_\alpha,j_{\alpha} k_{\alpha} a_{\alpha}} &= T_{-p_\beta,j_{\beta} k_{\beta} a_{\beta}} \notag \\
&= T_{-p,jka} - T_{-p,kja} \label{eq:T_jka2}
\end{align}
where our notation for the $\mathbf{M}_-$ and $\mathbf{T}_-$ matrix elements matches that of Ref.\@ \citenum{Banerjee:2019p224112}.  

To obtain working equations for $\mathbf{M}_-$ and $\mathbf{T}_-$ in IP-RADC, we write their spin-adapted matrix elements in terms of the spin-orbital Fock matrix eigenvalues ($\varepsilon_{p_\sigma}$), antisymmetrized two-electron integrals ($\braket{p_{\sigma_1}q_{\sigma_2}||r_{\sigma_3}s_{\sigma_4}}$), and $k$-th-order amplitudes of the effective Hamiltonian ($t_{i_{\sigma_1}}^{a_{\sigma_2}(k)}$ and  $t_{i_{\sigma_1}j_{\sigma_2}}^{a_{\sigma_3}b_{\sigma_4}(k)}$, \cref{eq: ex_op}), and perform spin-integration as follows:
\begin{align}
\varepsilon_{p_\alpha} &= \varepsilon_{p_\beta} = \varepsilon_{p} \label{eq:eps_spin_adapted}\\
\langle p_{\alpha} q_{\beta} || r_{\alpha} s_{\beta} \rangle &= \langle p_{\beta} q_{\alpha} || r_{\beta} s_{\alpha} \rangle = (pr | qs)\label{eq:mo_spin_adpated_abab}\\
\langle p_{\alpha} q_{\beta} || r_{\beta} s_{\alpha} \rangle &= \langle p_{\beta} q_{\alpha} || r_{\alpha} s_{\beta} \rangle = -(ps | qr)\label{eq:mo_spin_adpated_abba}\\
\langle p_{\alpha} q_ {\alpha} || r_{\alpha} s_{\alpha} \rangle &= \langle p_{\beta} q_ {\beta} || r_{\beta} s_{\beta} \rangle = (pr | qs) - (ps | qr)\label{eq : mo_spin_adpated_aaaa}\\
 t_{i_\alpha}^{a_\alpha(k)} &= t_{i_\beta}^{a_\beta(k)} = t_{i}^{a(k)}  \label{eq : t_ia}\\
 t_{i_{\alpha} j_{\beta}}^{a_{\alpha} b_{\beta}(k)} &=  t_{i_{\beta} j_{\alpha}}^{a_{\beta} b_{\alpha}(k)} = t_{ij}^{ab(k)} \label{eq : t_ijab_ab}\\
  t_{i_{\alpha} j_{\beta}}^{a_{\beta} b_{\alpha}(k)} &= t_{i_{\beta} j_{\alpha}}^{a_{\alpha} b_{\beta}(k)} = -t_{ij}^{ba(k)} \label{eq : t_ijab_abba}\\
t_{i_{\alpha} j_{\alpha}}^{a_{\alpha} b_{\alpha}(k)} &=  t_{i_{\beta} j_{\beta}}^{a_{\beta} b_{\beta}(k)}= t_{ij}^{ab(k)} - t_{ij}^{ba(k)}\label{eq : t_aaaa}
\end{align}
where $(pq | rs)$ denotes a Chemists' notation two-electron integral in the RHF MO basis. The resulting equations are used to solve the IP-RADC eigenvalue problem by computing the matrix-vector products
\begin{align}
\sigma_{-i} 
 &= \sum_{j} M_{-i, j} Y_{-j } + 2 \sum_{jka} M_{-i, jka}Y_{-jka} \notag \\
 &- \sum_{jka}  M_{-i, jka} Y_{-kja} \label{eq : sigma_i_adapted_3}\\
\sigma_{-jka} &=  \sum_{i} M_{-i, jka} Y_{-i} 
+  \sum_{ilb} M_{-jka, ilb} Y_{-ilb} \label{eq : sigma_jka_bab}
\end{align}
and iteratively optimizing the spin-adapted eigenvectors with elements $\mathbf{Y_-} = \{Y_{-i},Y_{-ija}\}$. The converged eigenvectors are then used to evaluate the spectroscopic factors as described in the Appendix. Expressions for $\mathbf{M}_+$, $\mathbf{T}_+$, and $\boldsymbol{\sigma}_+$ in EA-ADC can be obtained by replacing the 1h ($i$) and 2h-1p ($ija$) excitation labels with those of the 1p ($a$) and 2p-1h ($abi$) excitations.

Spin adaptation of the EA/IP-ADC equations provides significant computational savings for all steps of the ADC algorithm outlined in \cref{sec:implementation:overview}, with a three-fold reduction in the formal computational prefactor, as shown in \cref{tab: scaling}. 
Most significant savings are achieved for the solution of the EA/IP-ADC eigenvalue problem, where the matrix-vector products $\boldsymbol{\sigma}_\pm$ are expressed in terms of a much smaller number of the spin-adapted $\mathbf{M}_\pm$ matrix elements (e.g., $M_{-i,j}$, $M_{-i, jka}$, and $M_{-ilb, jka}$ in \cref{eq:M_ij,eq:M_ijka1,eq:M_ijka2,eq:M_ilbjka1,eq:M_ilbjka2}) compared to that of the spin-orbital implementation. We will analyze the computational efficiency of our EA/IP-RADC implementation in \cref{sec:results:accuracy_df}.

\subsection{Density fitting in EA/IP-ADC}
\label{sec:implementation:density-fitting}

In addition to using the conventional two-electron integrals ($(pq | rs)$ or $\braket{p_{\sigma_1}q_{\sigma_2}|r_{\sigma_3}s_{\sigma_4}}$), our EA/IP-RADC and UADC implementations feature support of the density fitting (DF) approximation\cite{Whitten:1973p4496,Dunlap:1979p3396,Vahtras:1993p514,Feyereisen:1993p359} that expresses the four-index integrals as a product of the two- and three-index tensors. For example, the two-electron integrals in EA/IP-RADC can be approximated as:
\begin{align}
	( pq | rs ) \approx \sum_{Q}^{N_{aux}} b_{pq}^{Q}b_{rs}^{Q} \label{eq:19}
\end{align}
where the $b_{pq}^{Q}$ coefficients are defined in terms of the MO's ($\phi_{p}$) and the auxiliary basis functions ($\chi_{P}$) as
\begin{align}
	 b_{pq}^{Q} &= \sum_{P}^{N_{aux}}( pq | P ) (J^{-\frac{1}{2}} )_{PQ}\label{eq: b_pq} \\
	 ( pq | P ) &=  \iint \phi_{p} (1) \phi_{q}(1) \frac{1}{r_{12}} \chi_{P}(2) dr_{1} dr_{2}\label{3_ind_int} \\
	 J_{PQ} &= \iint \chi_{P} (1) \frac{1}{r_{12}}\chi_{Q}(2)dr_{1} dr_{2}\label{eq: j_PQ} 
\end{align}
The use of the DF approximation has virtually no effect on the results of the EA/IP-ADC calculations (see \cref{sec:results:accuracy_df} for details) while offering two significant advantages: i) it greatly reduces the disk usage and the input/output operation count by avoiding the storage of large four-index tensors on disk, and ii) it dramatically lowers the computational scaling of the two-electron integral transformation from $\mathcal{O}(N^5)$ to $\mathcal{O}(N_{aux}N^3)$  (where $N$ and $N_{aux}$ are the numbers of  $\phi_{p}$ and $\chi_{P}$, respectively), providing significant computational savings for large molecules and basis sets. In our implementation, we store the $b_{pq}^{Q}$ tensors either in memory or on disk and recompute sectors of $( pq | rs )$ on demand. Similar to other methods that employ the DF approximation,\cite{Werner:2003p8149,Bozkaya:2014er,Epifanovsky:2013gd,Wang:2016p4833} our density-fitted EA/IP-ADC methods support using the JKFIT and RI auxiliary basis sets. The RI basis functions are used to approximate two-electron integrals in the EA/IP-ADC calculations, while the JKFIT auxiliary basis set optimized for fitting the Coulomb and exchange integrals in the mean-field (SCF) computations is used in the reference Hartree--Fock calculation. We note that the DF approximation has been previously used in other efficient single-reference ADC implementations.\cite{Helmich:2014p35,Mester:2018p094111,Herbst:2020pe1462,Liu:2020p174109}

\section{Computational details}
\label{sec:comp_details}

All EA/IP-ADC($n$) calculations ($n \le 3$) were performed using the ADC module in \textsc{PySCF}\cite{Sun:2020p024109,Banerjee:2019p224112} and 
used the singly- or doubly-augmented Dunning's correlation consistent basis sets denoted as aug-cc-pV$X$Z or d-aug-cc-pV$X$Z, respectively.\cite{Dunning:1989p1007,Kendall:1992p6796,Woon:1994p2975} Specifically, in \cref{sec:results:accuracy_df}, we used aug-cc-pVQZ to benchmark the accuracy of the density fitting approximation and aug-cc-pVDZ to test computational efficiency of our implementation. All ADC(2) and MP2 computations of the TEMPO radical (\cref{sec:results:TEMPO}) and DNA base pairs (\cref{sec:results:base_pairs}) employed the aug-cc-pVTZ basis sets. For the ADC(3) and MP3 calculations we used a modified aug-cc-pVTZ basis set where the diffuse functions were removed from the hydrogen atoms. We denote this basis set as aug(nH)-cc-pVTZ. Additionally, for the DNA base pairs, we employed the doubly-augmented d-aug-cc-pVDZ basis set.

Geometries of molecules in \cref{sec:results:accuracy_df} were optimized using coupled cluster theory with single, double, and perturbative triple excitations (CCSD(T)).\cite{Raghavachari:1989p479,Bartlett:1990p513} 
The EA/IP-ADC($n$) calculations ($n\le 3$) of the TEMPO radical and anion used structures optimized using the $n$-th-order M\o ller-Plesset perturbation theory\cite{Moller:1934kp618} (MP$n$) and the same basis set. The EA-ADC($n$) calculations for the base pairs and their corresponding anions were performed with the MP2/aug-cc-pVTZ optimized structures. 
All geometry optimizations were performed using the \textsc{PSI4} program\cite{Smith:2020p184108} and the optimized structures are reported in the Supporting Information. All EA/IP-ADC($n$) and MP$n$ calculations employed the density fitting approximation. The correlation MP$n$ energies and ADC($n$) excitation energies were computed using the resolution-of-identity (aug-cc-pV$X$Z-RI) auxiliary basis sets, while the SCF energies and molecular orbitals were obtained using the aug-cc-pV$X$Z-JKFIT auxiliary basis sets,\cite{Weigend:2002p4285,Weigend:2002p3175,Hattig:2005p59} where $X$ corresponded to that of the main one-electron basis set.

Throughout the manuscript, positive electron affinity implies exothermic electron attachment (i.e., EA = $E_{N}$ $-$ $E_{N+1}$), whereas a positive ionization energy corresponds to an endothermic process (IP = $E_{N-1}$ $-$ $E_{N}$). While the vertical electron affinities (VEA's) were obtained directly from our EA-ADC($n$) calculations, the corresponding adiabatic electron attachment (AEA's) energies were determined as follows:
\begin{align}
      \label{eq:aea_from_vea}
      AEA = VEA + \Delta E_{MPn}  
\end{align}
where $ \Delta E_{MPn} = E^{N+1}_{MPn}$(neutral) $-$ $E^{N+1}_{MPn}$(anion) is the difference of the MP$n$ energies of the anionic system computed at the neutral and anion optimized geometries. 
The TEMPO radical AEA's have been corrected to account for the zero-point vibrational energy (ZPVE)
\begin{align}
      AEA(ZPVE) &= AEA \notag \\
      &+ ZPVE(N) - ZPVE(N+1)
\end{align}
where $ZPVE(N)$ and $ZPVE(N+1)$ are the ZPVE's of the neutral and anionic TEMPO, respectively, computed at the MP2/aug-cc-pVDZ level of theory.

The nature of the electron-attached states was analyzed by visualizing the Dyson orbitals,\cite{Melania:2007p234106} which are computed from the overlap of the ground state wavefunction and the electron-attached/ionized states:
\begin{align}
	|\phi^{Dyson}_{\pm\mu}\rangle =  \sum_{p} X_{\pm p\mu}  |\phi_{p}\rangle \label{eq:19}
\end{align}
where $|\phi_{p}\rangle$ are the reference Hartree--Fock orbitals and $X_{\pm p\mu}$ are the EA/IP-ADC spectroscopic amplitudes defined in \cref{eq: spec_amp}. All Dyson orbitals were computed using the aug-cc-pVDZ basis set. 

\begin{figure*}[t!]
    \includegraphics[width=5.5in]{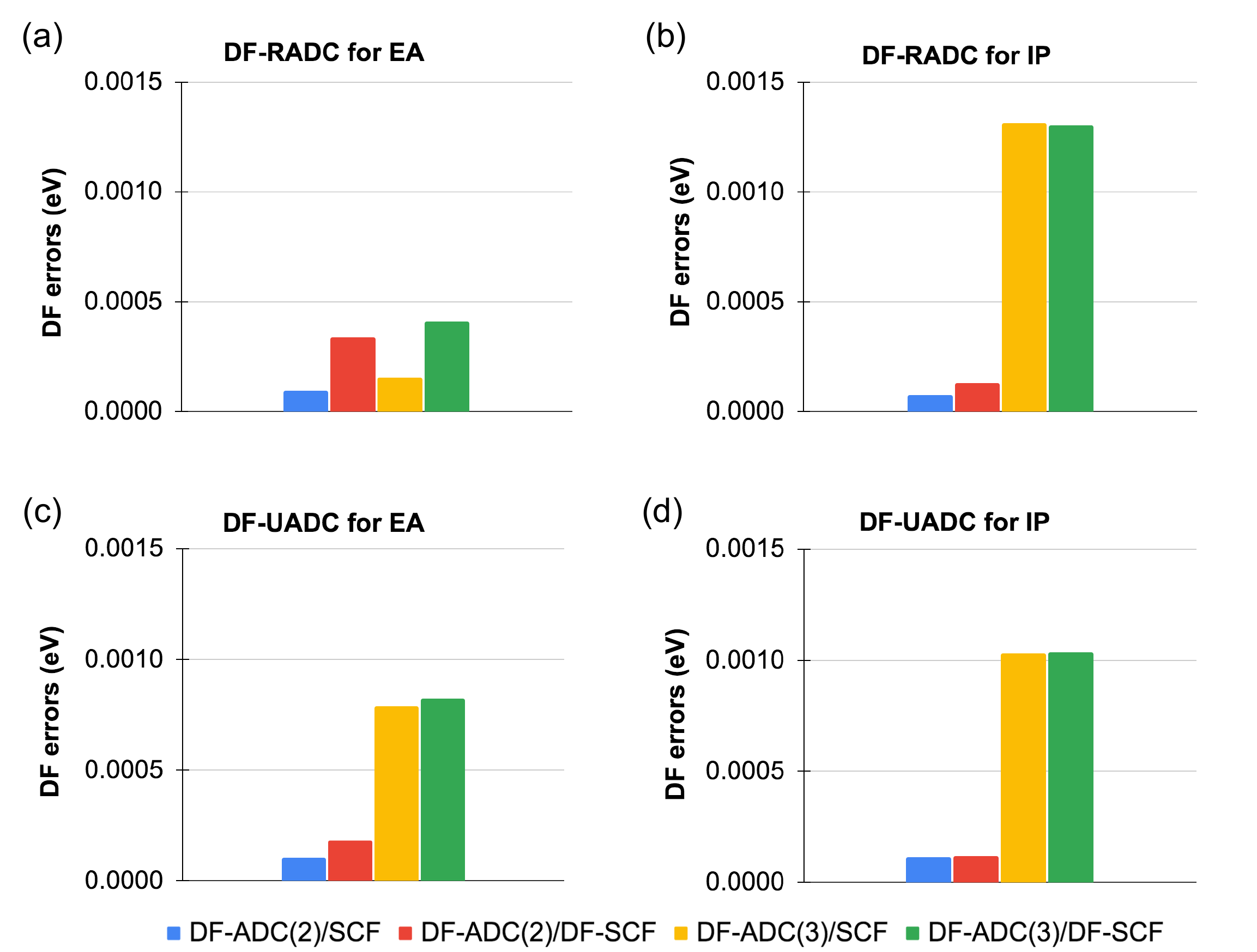}
	\captionsetup{justification=raggedright,singlelinecheck=false,font=footnotesize}
	\caption{Mean absolute errors (eV) of the density fitting approximation in the vertical electron attachment or ionization energies for EA-RADC($n$) (a), IP-RADC($n$) (b), EA-UADC($n$) (c), and IP-UADC($n$) (d) with $n = 2, 3$ (aug-cc-pVQZ basis set). For each method, two approximations are considered, with (DF-ADC($n$)/DF-SCF) and without (DF-ADC($n$)/SCF) density fitting in the reference SCF calculation. The RADC($n$) and UADC($n$) results were obtained for sets of five closed-shell and five open-shell molecules, respectively. See Tables S1 and S2 in the Supporting Information for data on individual molecules.}
	\label{fig:df-accuracy}
\end{figure*}

The TEMPO radical photoelectron spectra were simulated by plotting the IP-ADC density of states
\begin{align}
	\label{eq: PES}
	A(\omega) &= -\frac{1}{\pi} \mathrm{Im} \left[ \sum_{\mu}\frac{P_{-\mu}}{\omega - \omega_{\mu} + i\eta} \right]
\end{align}
for a range of frequencies $\omega$, where $\omega_{\mu}$ denotes the ionization energies obtained by solving \cref{eq: eig}, $\eta$ is a small broadening, and $P_{-\mu}$ are the spectroscopic factors computed using \cref{eq:spec_factors}. The simulated spectra were compared to the experimental photoelectron spectrum that was digitized using the WebPlotDigitizer program.\cite{Rohatgi2020}

\section{Results}
\label{sec:results}

\subsection{Accuracy of the density fitting approximation and analysis of the computational efficiency}
\label{sec:results:accuracy_df}

In this section, we benchmark the accuracy of the density fitting (DF) approximation for EA/IP-RADC($n$) ($n$ = 2, 3) and analyze the efficiency of our restricted (RADC) and unrestricted (UADC) implementations. \cref{fig:df-accuracy} shows the errors in the vertical electron attachment (EA) and ionization (IP) energies computed using the density-fitted EA- or IP-ADC($n$) methods, respectively, relative to those of the conventional EA/IP-ADC($n$) ($n$ = 2, 3) methods without the density fitting approximation. For each EA/IP-DF-ADC($n$) method, results are presented for sets of five closed-shell (\ce{H2O}, \ce{HF}, \ce{N2}, \ce{SiH2}, \ce{H2CO}) and five open-shell (\ce{OH}, \ce{NH}, \ce{O2}, \ce{NH2}, \ce{CH3}) molecules computed using the RADC and UADC implementations, respectively (with data for individual molecules available in Tables S1 and S2 of the Supporting Information). Furthermore, we consider two types of the DF approximations, with and without density fitting in the reference SCF calculation, denoted as DF-ADC($n$)/DF-SCF and DF-ADC($n$)/SCF, respectively.

For all methods considered, the errors of the DF approximation are very small and do not exceed 0.0021 eV. The largest mean absolute errors (MAE's) of $\sim$ 0.001 eV are observed for IP-DF-ADC(3) of closed- and open-shell molecules and for EA-DF-ADC(3) of open-shell molecules. The EA/IP-DF-ADC(2) methods consistently produce smaller errors with MAE's less than 0.0005 eV. Incorporating the DF approximation in the SCF reference calculations increases the computed MAE's by no more than 0.0003 eV. Overall, we find that the DF approximation in EA/IP-DF-ADC($n$) ($n$ = 2, 3) is very accurate, which is consistent with findings for other quantum chemical methods when computing relative energies.\cite{Epifanovsky:2013gd,Bozkaya:2014p2371,Wang:2016p4833,Bozkaya:2016p1179,Hattig:2000p5154} 

\begin{figure*}[t!]
    \includegraphics[width=6.7in]{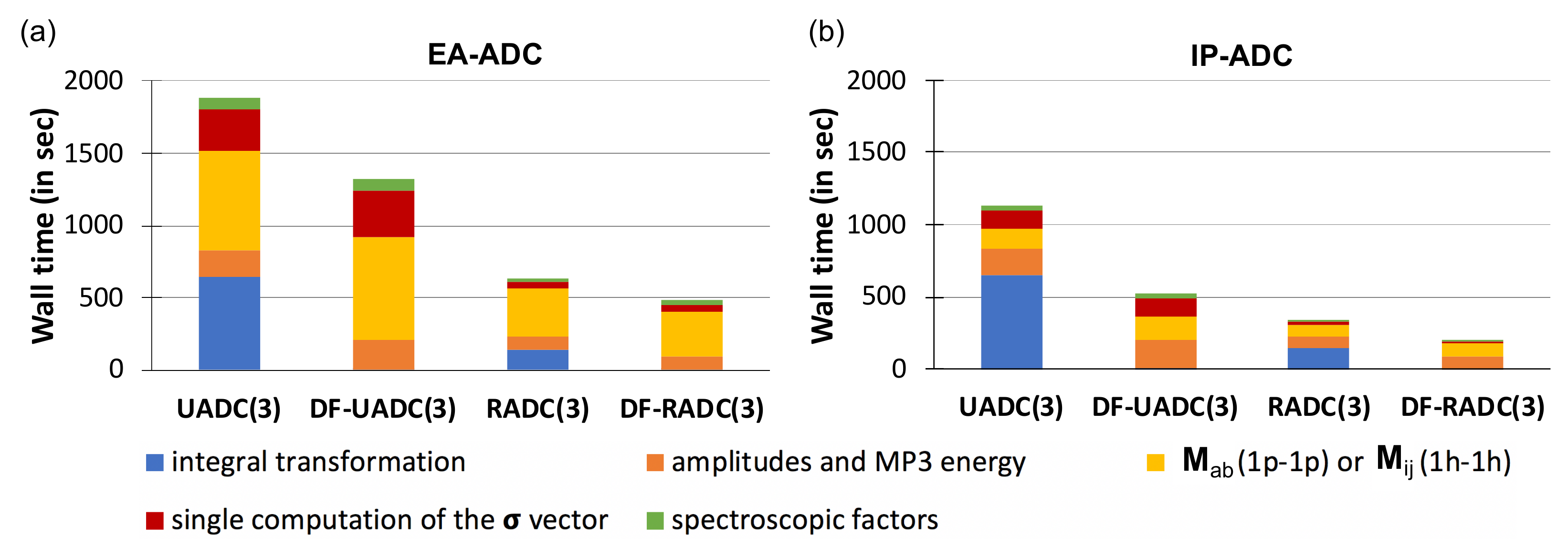}\label{fig:timings}
	\captionsetup{justification=raggedright,singlelinecheck=false,font=footnotesize}
	\caption{Comparison of the computational wall time (in sec) for the restricted (R) and unrestricted (U) implementations of the conventional (out-of-core) and density-fitted (DF) ADC algorithms for EA-ADC(3) (a) and IP-ADC(3) (b). Data is shown for the cytosine molecule with the aug-cc-pVDZ basis (229 basis functions, 3 lowest EA or IP). Wall time for each step of the ADC algorithm outlined in \cref{sec:implementation:overview} is shown with a different color. All calculations were performed on a single computer with a 12-core Intel Xeon Gold 6136 3.00GHz CPU.}
	\label{fig:timings}
\end{figure*}   

We now analyze the efficiency of our EA/IP-RADC(3) implementations by comparing the computational wall time for calculating the three lowest-energy EA or IP of the cytosine molecule with the aug-cc-pVDZ basis set (229 basis functions). We employ four algorithms for each method: i) out-of-core UADC(3), ii) density-fitted UADC(3), iii) out-of-core RADC(3), and iv) density-fitted RADC(3). The computed wall time is broken down into contributions from each step of the ADC algorithm outlined in \cref{sec:implementation:overview}.

Taking advantage of the spin adaptation in the RADC algorithm reduces the total time of the EA- and IP-ADC(3) calculations by approximately a factor of three. Significant savings are observed for all steps of the ADC computation, with $\sim$ 2.5-fold speedup for calculating the amplitudes of the effective Hamiltonian and the 1p-1p/1h-1h blocks of the $\mathbf{M}_\pm$ matrix and more than four-fold reduction in time for calculating the $\boldsymbol{\sigma}_\pm$ vector and spectroscopic factors, in a good agreement with theoretical prefactors reported in \cref{tab: scaling}. Introducing the DF approximation drastically lowers the computational cost of the two-electron integral transformation, which has a substantial effect on the wall time of the EA/IP-UADC(3) calculations resulting in a 1.5- to 2-fold speedup. Timings for other steps of the ADC algorithm are not affected by the density fitting. Combining the spin adaptation and density fitting allows to achieve nearly a  four-fold speedup for both EA- and IP-ADC(3), pushing the limits of their applicability to larger chemical systems and basis sets. In \cref{sec:results:TEMPO,sec:results:base_pairs}, we demonstrate this by applying EA/IP-RADC($n$) ($n$ = 2, 3) to the TEMPO radical and the DNA base pairs.

\subsection{TEMPO radical}
\label{sec:results:TEMPO}

\begin{figure}[t!]
         \includegraphics[width=1.5in]{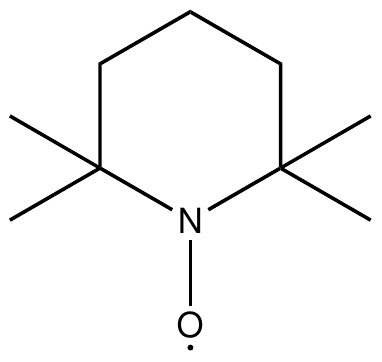}
	\captionsetup{justification=raggedright,singlelinecheck=false,font=footnotesize}
	\caption{Molecular structure of the TEMPO radical.}
	\label{fig:tempo_structure}
\end{figure} 

We first study the (2,2,6,6-tetramethylpiperidin-1-yl)oxyl radical, commonly known as TEMPO (\cref{fig:tempo_structure}), that is widely used as a catalyst in organic synthesis, as a mediator in polymerization reactions, and as a structural probe for performing electron spin resonance spectroscopy on biological systems.\cite{Barriga:2001p0563,Galli:2008p705,Likhtenshtein:2008nitroxides} TEMPO and its derivatives are also used in solar cells, lithium batteries, and radical sensors.\cite{Nakahara:2011p227,Janoschka:2015p78,Nakahara:2006p921,Blinco:2008p6763,Kato:2010p464,Buhrmester:2006pa1800} 
 
 \begin{figure*}[t!]
     \subfigure[]{\includegraphics[width=2.25in]{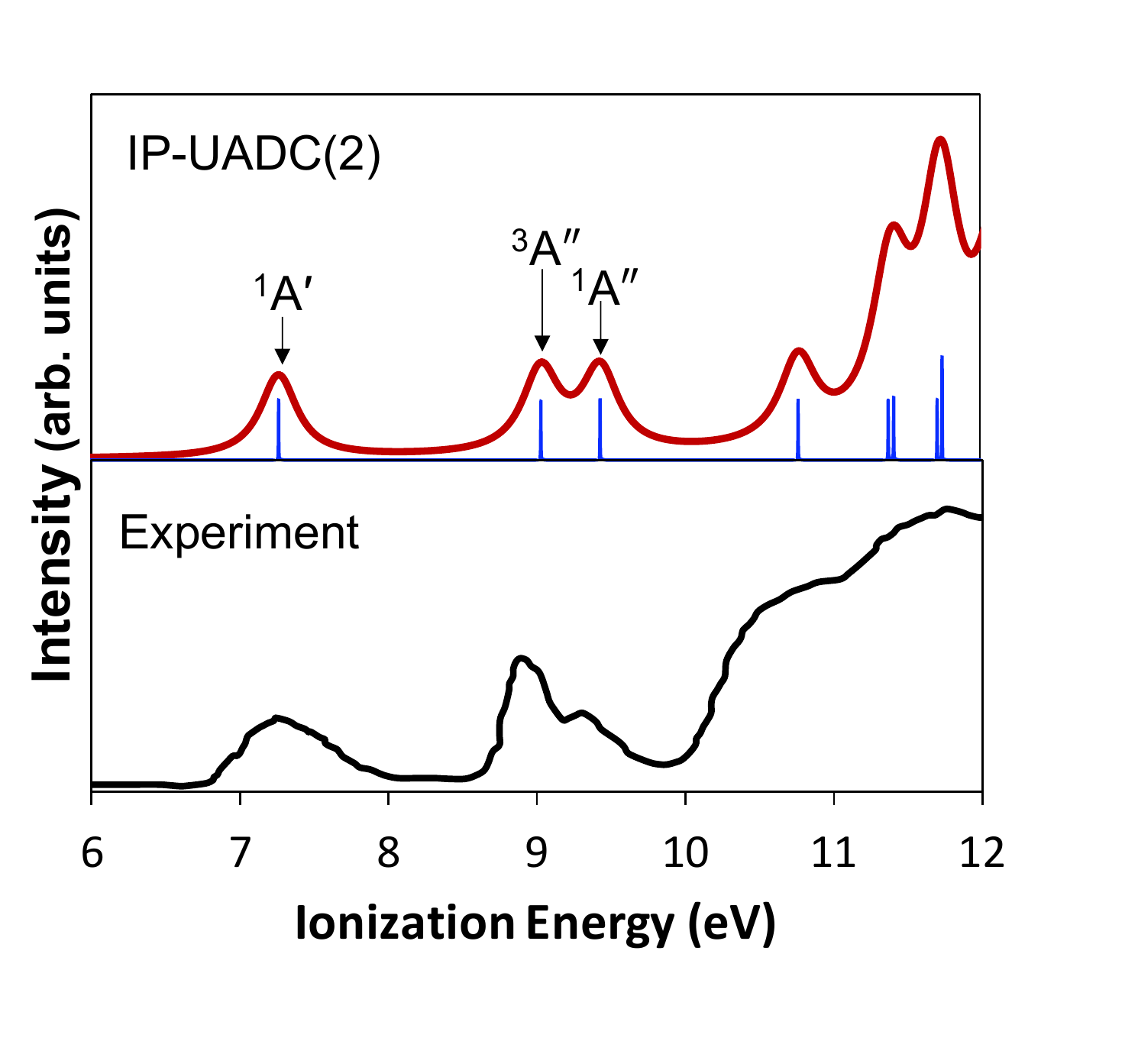}\label{fig:adc2_spectra}}  \qquad
     \subfigure[]{\includegraphics[width=2.25in]{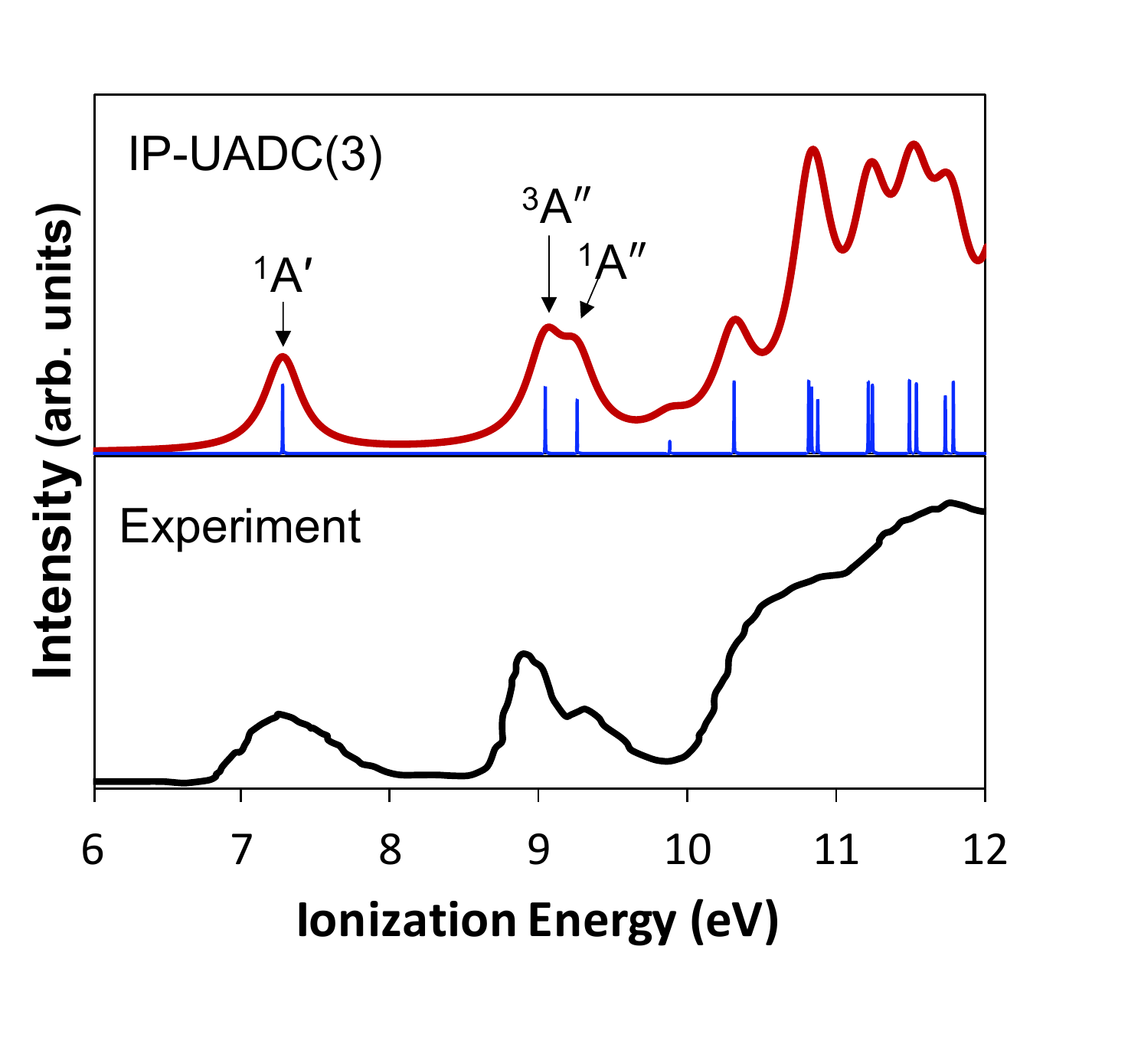}\label{fig:adc3_spectra}}  
	\captionsetup{justification=raggedright,singlelinecheck=false,font=footnotesize}
	\caption{Density of states of the TEMPO radical computed using IP-UADC(2)/aug-cc-pVTZ (a) and IP-UADC(3)/aug(nH)-cc-pVTZ (b) with a broadening of 0.15 eV, compared to the experimental photoelectron spectrum from Ref.\@ \citenum{Kubala:2013p2033}. Densities of states were shifted by 1.02 ($n$ = 2) and $-$0.3 eV ($n$ = 3) to reproduce the position of the first peak in the experimental spectrum. 
	}
	\label{fig:tempo_dos_photoelectron}
\end{figure*}

\begin{figure*}[t!]
     \subfigure[]{\includegraphics[width=2.5in]{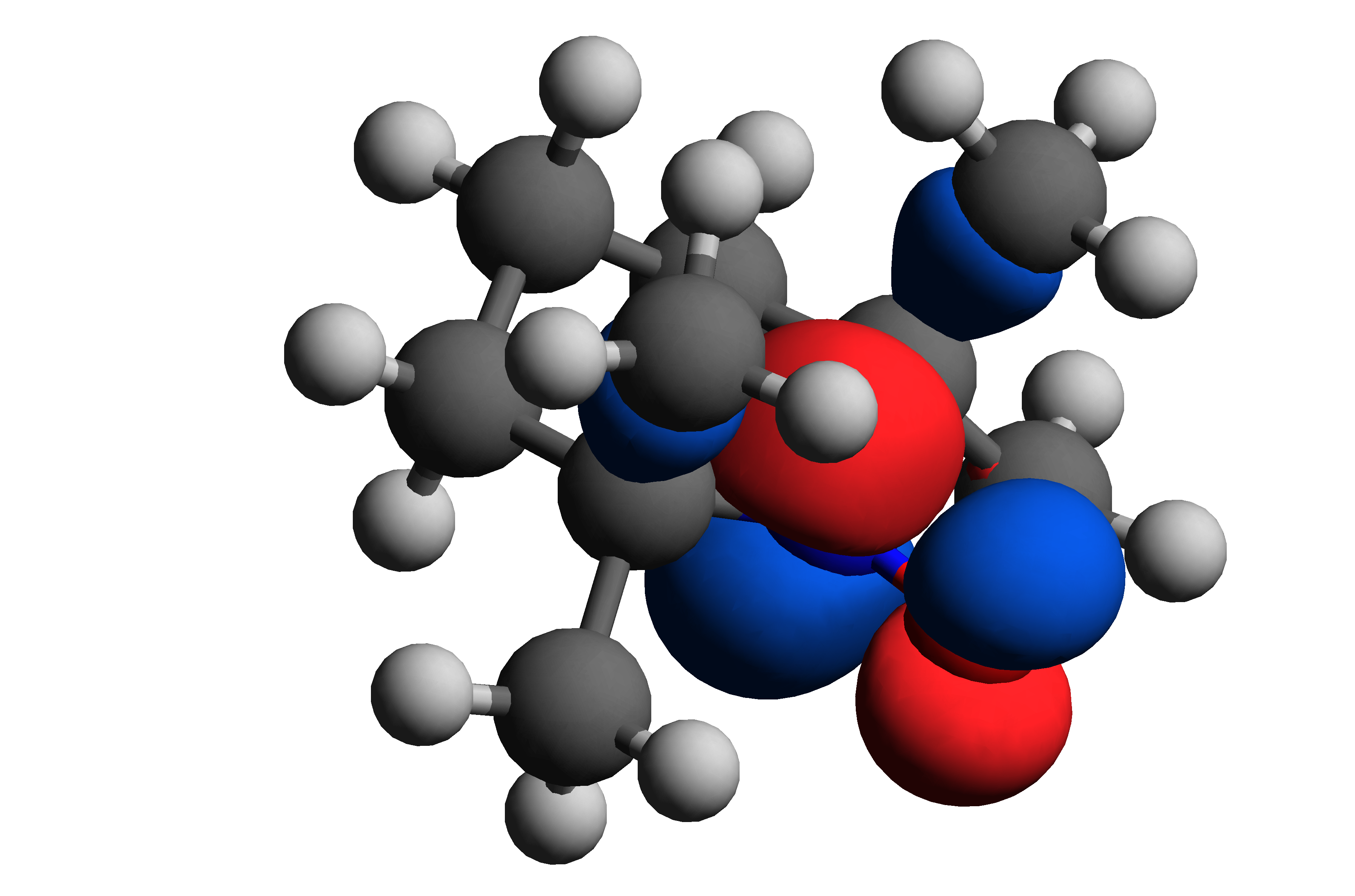}\label{fig:singlet_a_prime}}  \qquad
     \subfigure[]{\includegraphics[width=2.5in]{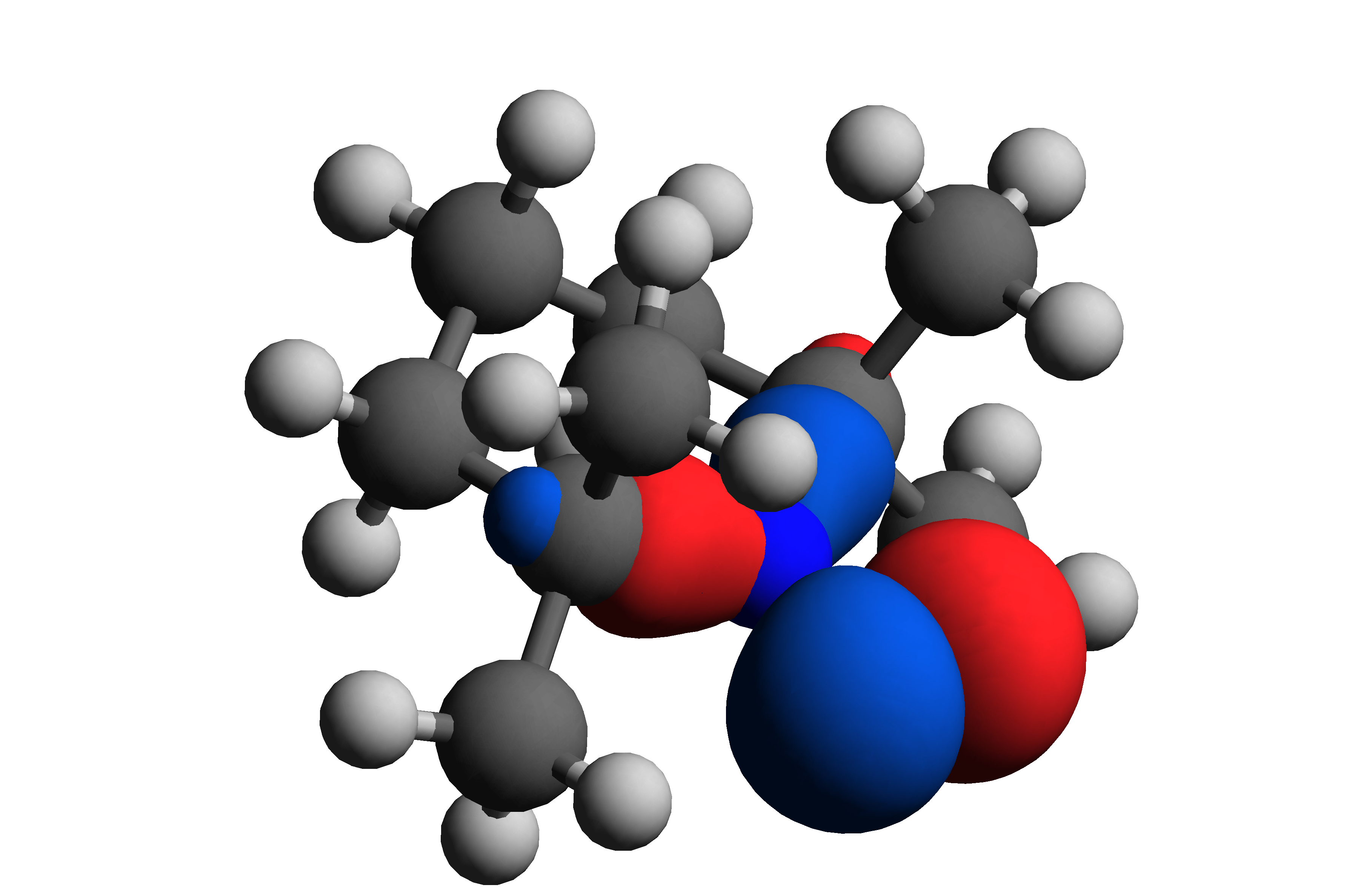}\label{fig:triplet_a_dprime}}  
     \subfigure[]{\includegraphics[width=2.5in]{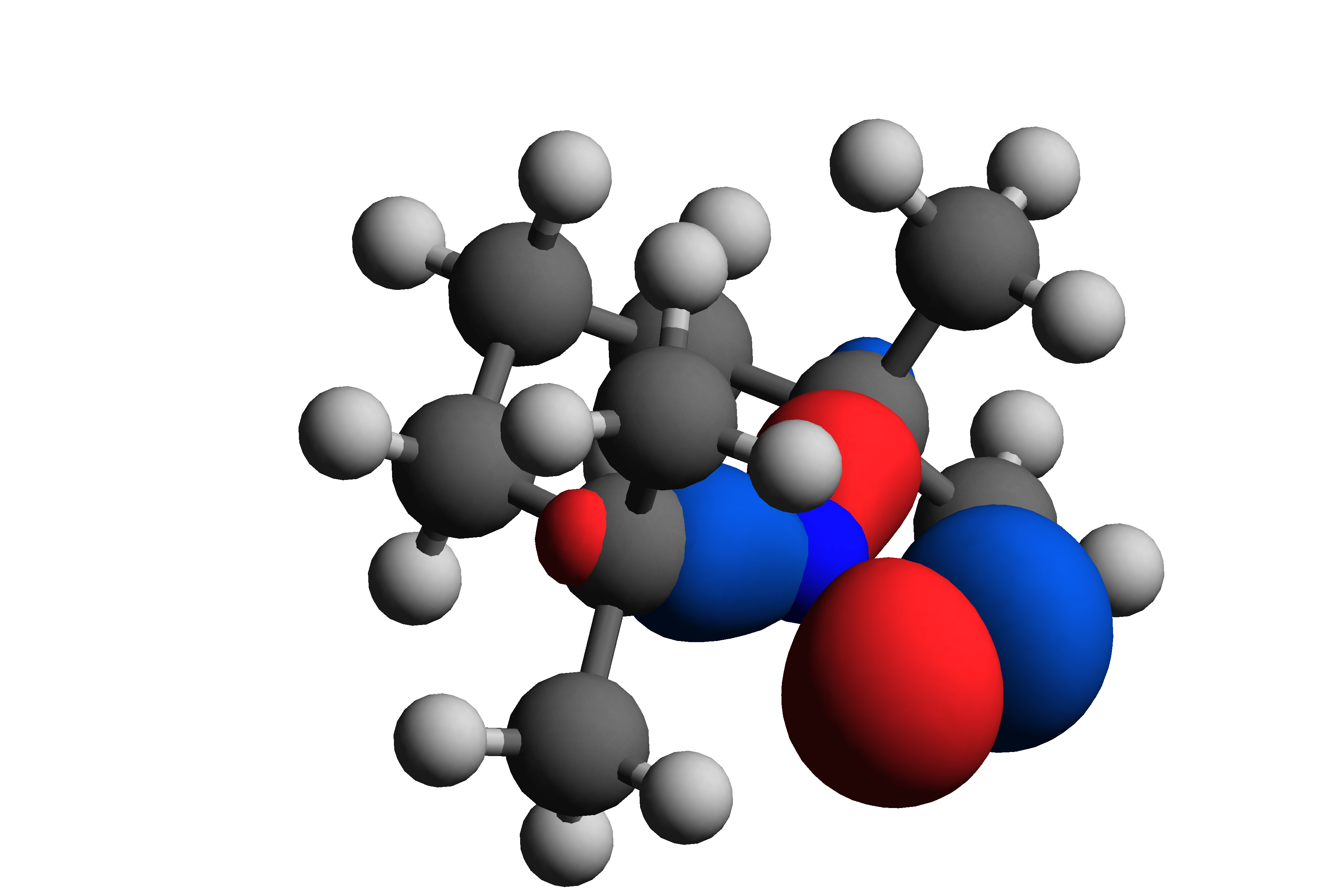}\label{fig:singlet_a_dprime}}
	\captionsetup{justification=raggedright,singlelinecheck=false,font=footnotesize}
	\caption{Dyson orbitals for to the three lowest-energy ionization transitions of the neutral TEMPO radical corresponding to the $1^{1}A'$ (a), $1^{3}A''$ (b) and  $1^{1}A''$ (c) states of the TEMPO cation computed using the IP-UADC(2) method with the aug-cc-pVTZ basis set.}
	\label{fig:tempo_ip_dyson}
\end{figure*}

The diverse applications of the TEMPO radical have motivated several experimental and theoretical investigations of its electronic structure. Photoionization of TEMPO has been thoroughly studied, experimentally and theoretically, by Ljubi\'c, Novak, and co-workers using UV/Vis and X-ray photoelectron spectroscopy.\cite{Novak:2004p7628,Kovavc:2014p10734,Ljubic:2014p2333,Ljubic:2016p10207} These studies demonstrated a good agreement between the experimental photoelectron spectra and theoretical spectra simulated using density functional theory (DFT). However, electron attachment to TEMPO has not been investigated as extensively as its photoionization. Experimentally, the TEMPO anion has been studied by Kubala et al.\@ using dissociative electron attachment spectroscopy, electron energy-loss spectroscopy, and vibrational excitation cross section measurements.\cite{Kubala:2013p2033} Their results indicated that the energy of the electron-attached state is either slightly below or slightly above that of the neutral TEMPO radical, but did not allow to distinguish between these two possibilities. These findings were supported by the DFT calculations (B3LYP/6-311++G(2d,p)) that predicted the endothermic vertical electron affinity (VEA) of $-$0.3 eV and exothermic zero-point-energy corrected adiabatic electron affinity (AEA) of 0.14 eV. In another theoretical study, Sohn et al.\@ used a combination of the B3LYP and MP2 methods with the 6-311++G(3df,2pd) basis set and computed AEA to be 0.64 eV.\cite{Sohn:2015p49} A more recent study by Gunasekara et al.\@ using coupled cluster theory with single, double, and perturbative triple excitations (CCSD(T)) and the aug-cc-pVTZ basis set reports AEA of 0.44 eV.\cite{Gunasekara:2019p1882} 

\begin{table*}[t!]
\begin{threeparttable}
	\captionsetup{justification=raggedright,singlelinecheck=false,font=footnotesize}
	\caption{Vertical (VEA) and adiabatic (AEA) electron attachment energies (in eV) of the TEMPO radical computed using MP$n$ and EA-UADC($n$) methods. 
Also shown are AEA calculated with the zero-point vibrational energy (ZPVE) correction and best available results from other theoretical methods.}
	\label{tab:TEMPO}
	\footnotesize
	\setstretch{1}
    \begin{tabular}{L{5.2cm}C{3.5cm}C{3.5cm}C{4cm}}
       \hline
        \hline
        \multicolumn{1}{c}{Method} &\multicolumn{1}{c}{VEA} &\multicolumn{1}{c}{AEA} &\multicolumn{1}{c}{AEA (ZPVE-corrected)} \\
               \hline
MP2/aug-cc-pVTZ                          		&$-$0.21 		&0.57		&0.70 \\
MP3/aug(nH)-cc-pVTZ                           		&$-$0.58		&0.23		&0.35 \\
EA-UADC(2)/aug-cc-pVTZ     			&$-$0.30		&0.49		&0.61 \\
EA-UADC(3)/aug(nH)-cc-pVTZ      	        		&$-$0.72		&0.08		&0.21 \\
Reference               &$-$0.3\tnote{a}          &0.64\tnote{b}, 0.44\tnote{c}\ 	& 0.14\tnote{a} \\
     		        \hline
        \hline
    \end{tabular}
    \begin{tablenotes}
    \item[a] B3LYP/6-311++G(2d,p) from Ref.\@ \citenum{Kubala:2013p2033}.
    \item[b] MP2/6-311++G(3df,2pd) from  Ref.\@ \citenum{Sohn:2015p49}.
    \item[c] CCSD(T)/aug-cc-pVTZ from  Ref.\@ \citenum{Gunasekara:2019p1882}.
     \end{tablenotes}
\end{threeparttable}
\end{table*}
\vspace{0.1cm}

To check the accuracy of our ADC methods for the TEMPO radical, we first compute its ionization spectrum using IP-UADC($n$) ($n$ = 2, 3) and compare it to the experimental photoelectron spectrum reported by Kubala et al.\cite{Kubala:2013p2033} \cref{fig:tempo_dos_photoelectron} shows the density of occupied states (DOS) of TEMPO computed using IP-UADC($n$) ($n$ = 2, 3) with a broadening of 0.15 eV. For each method, DOS was shifted to align the first IP with the maximum of the first peak in the experimental spectrum. The IP-UADC(3) DOS required a smaller shift ($-0.3$ eV) compared to that of IP-UADC(2) (1.02 eV), indicating a smaller error of the third-order approximation in the computed IP's, due to its higher-level description of electron correlation effects. The shifted DOS computed using both IP-UADC(2) and IP-UADC(3) agree well with the experimental photoelectron spectrum, accurately reproducing peak spacings and relative peak intensities, with IP-UADC(3) showing a better agreement than IP-ADC(2) for ionization energies greater than 10 eV. Upon shifting, both methods predict a signal at $\sim$ 7.2 eV, corresponding to the $X^2A'\rightarrow1^1A'$ photoelectron transition, and two peaks at $\sim$ 9.0 and 9.3 eV that can be assigned as the $X^2A'\rightarrow1^3A''$ and $X^2A'\rightarrow1^1A''$ transitions, respectively. \cref{fig:tempo_ip_dyson} shows the Dyson orbitals for each of these photoionization transitions computed using  IP-UADC(2). All three Dyson orbitals are localized on the NO group of the TEMPO radical (\cref{fig:tempo_structure}), either in or out of the plane of its six-membered ring. The out-of-plane orbital shows a greater localization on the N atom, while the in-plane orbitals are more localized on the O atom. While IP-UADC(2) predicts both $X^2A'\rightarrow1^3A''$ and $X^2A'\rightarrow1^1A''$ transitions to have similar spectral density, IP-UADC(3) yields a smaller spectral density for the higher-energy $X^2A'\rightarrow1^1A''$ transition, in a good agreement with the experiment. The IP-UADC(3) DOS also shows a weak satellite transition at $\sim$ 10 eV, which is not visible in the experimental photoelectron spectrum.

We now turn our attention to VEA and AEA of the TEMPO radical computed using the MP$n$ and EA-UADC($n$) methods ($n$ = 2, 3) and presented in \cref{tab:TEMPO}. In agreement with a previous DFT study,\cite{Kubala:2013p2033} all MP$n$ and EA-UADC($n$) methods predict a negative VEA ranging between $-$0.72 eV (EA-UADC(3)) to $-$0.21 eV (MP2), indicating an endothermic vertical electron attachment. 
The VEA computed using MP$n$ and EA-UADC($n$) are close to each other at each order in perturbation theory (within 0.15 eV) and become more negative as the order increases from $n$ = 2 to 3. 

Upon structural relaxation, the electron-attached state of the TEMPO radical lowers its energy below that of the neutral radical and the electron attachment becomes energetically favorable, as evidenced by the positive AEA computed by all four MP$n$ and EA-UADC($n$) approximations. Out of all methods, the AEA of EA-UADC(2) (0.49 eV) shows the best agreement with the AEA from the CCSD(T) study by Gunasekara and co-workers (0.44 eV),\cite{Gunasekara:2019p1882} which used the same aug-cc-pVTZ basis set, but a different (DFT-optimized) geometry. The EA-UADC(3) method predicts a small positive AEA of 0.08 eV, in a good agreement with experimental and theoretical findings of Kubala et al.\cite{Kubala:2013p2033} \cref{tab:TEMPO} also reports AEA corrected by the zero-point vibrational energies (ZPVE) of the neutral and anionic states (see \cref{sec:comp_details} for details). Including the zero-point vibrational effects increases the EA-UADC(3) AEA to 0.21 eV, which agrees well with the experimental results.

\begin{figure*}[t!]
     \subfigure[]{\includegraphics[width=2in]{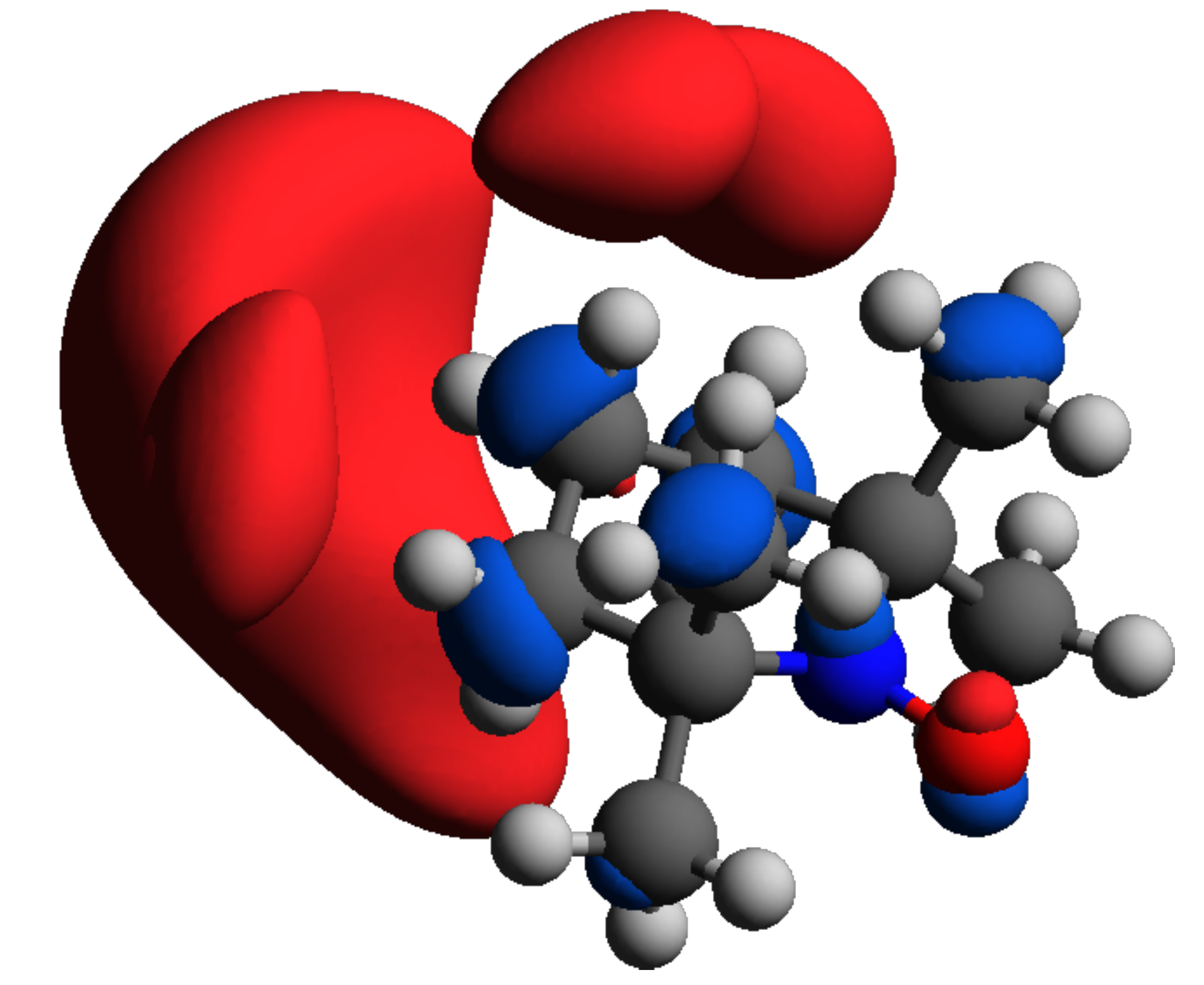}\label{fig:tempo_dyson_ea_neutral}} \qquad
      \subfigure[]{\includegraphics[width=2in]{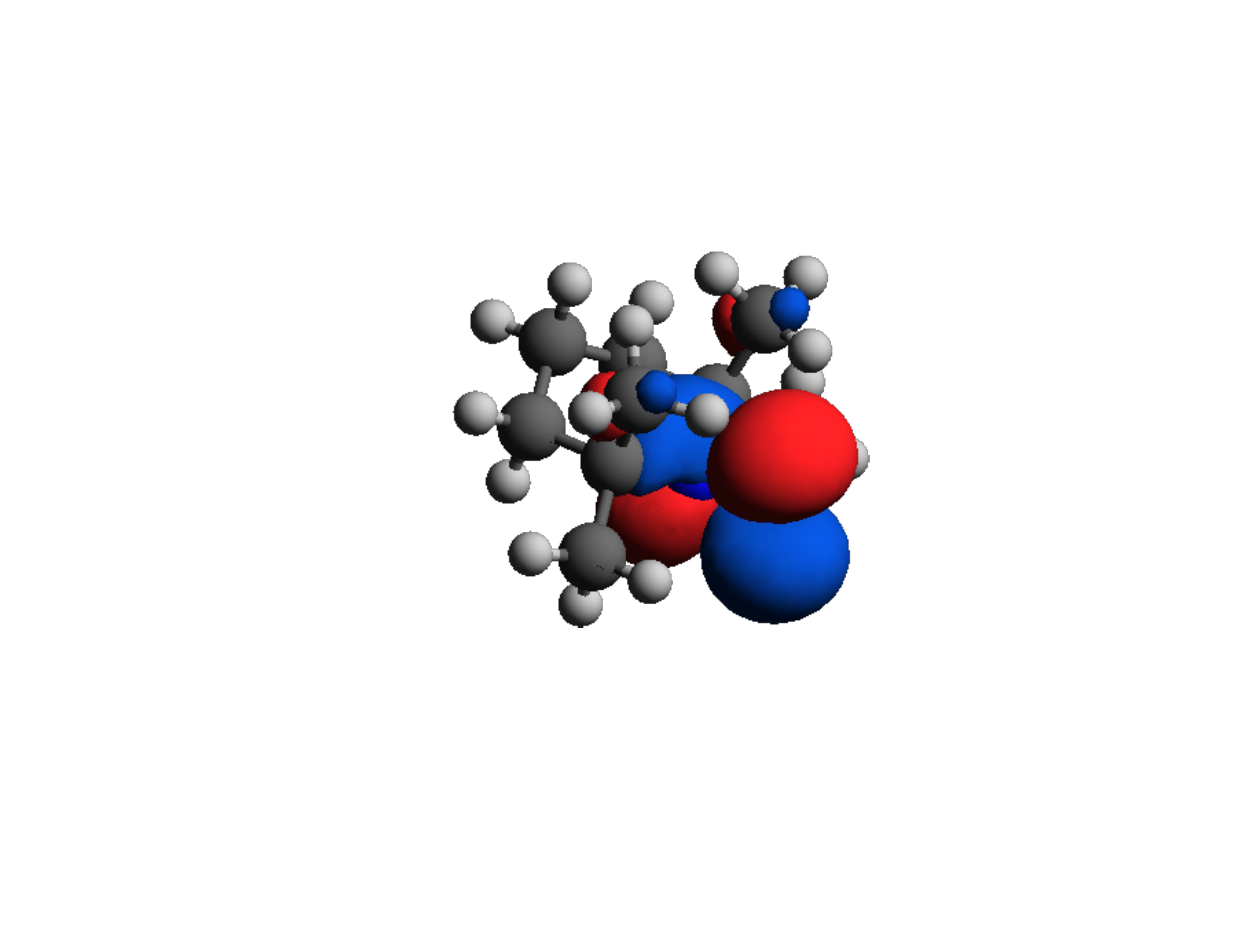}\label{fig:tempo_dyson_ea_anion}}  
	\captionsetup{justification=raggedright,singlelinecheck=false,font=footnotesize}
	\caption{Dyson orbitals corresponding to the lowest-energy electron attachment to the TEMPO radical at the neutral (a) and anion (b) equilibrium geometries computed using EA-UADC(3) with the aug-cc-pVDZ basis set.}
	\label{fig:tempo_dyson}
\end{figure*} 

\cref{fig:tempo_dyson_ea_neutral,fig:tempo_dyson_ea_anion} show the EA-UADC(3) Dyson orbitals of the TEMPO radical computed at the equilibrium geometries of the neutral and electron-attached states. At the neutral equilibrium geometry, the Dyson orbital is localized outside of the molecule, indicating a dipole-bound nature of the electron-attached state (\cref{fig:tempo_dyson_ea_neutral}). Once the TEMPO geometry is relaxed to that of its anion, the lowest-energy electron-attached state acquires the valence-bound character (\cref{fig:tempo_dyson_ea_anion}) where most of the orbital is localized on the NO group of the radical.

\subsection{DNA base pairs}
\label{sec:results:base_pairs}

\begin{figure*}[t!]
     \subfigure[]{\includegraphics[width=3.0in]{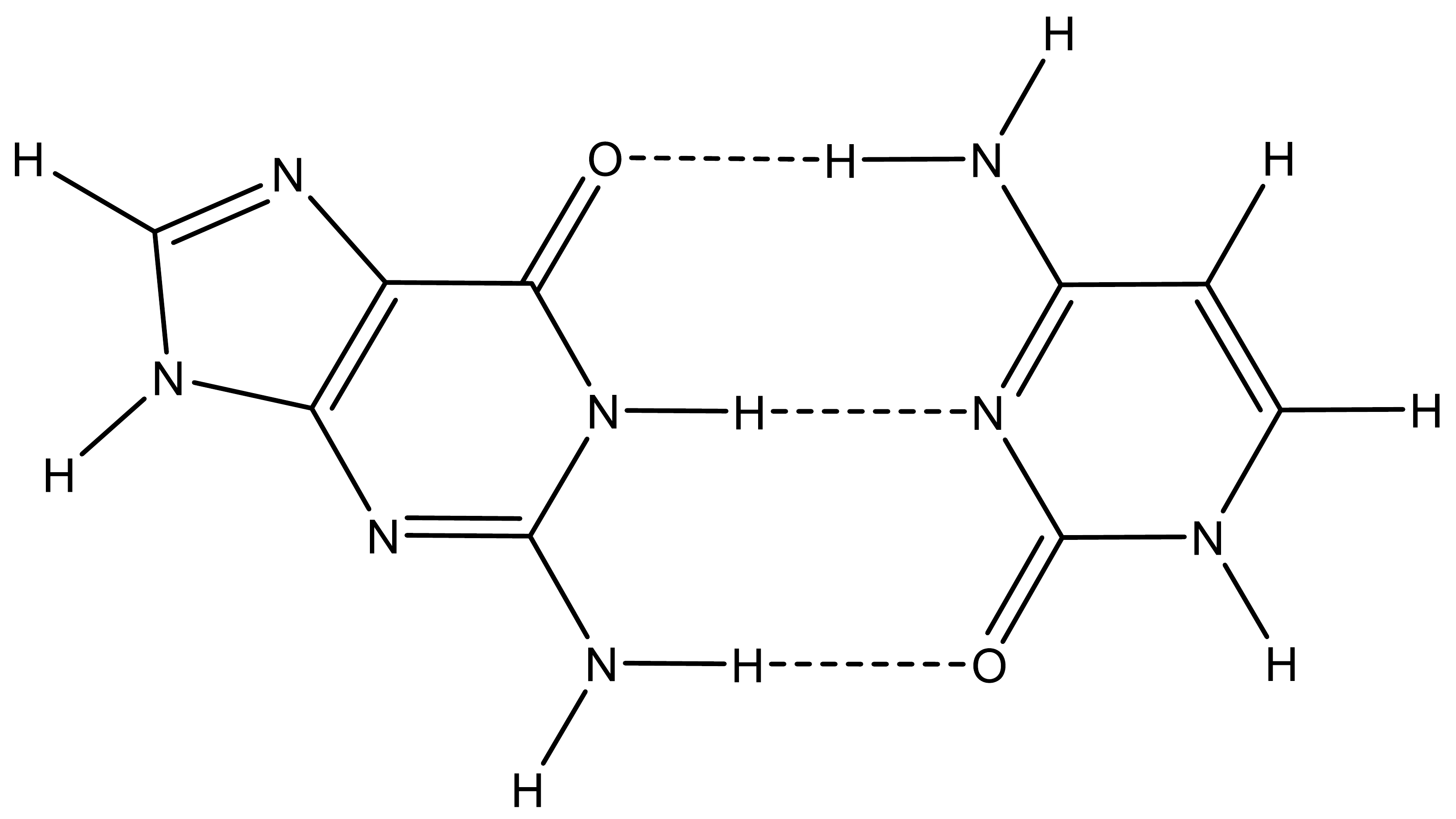}\label{fig:gc_structure}}  \qquad
     \subfigure[]{\includegraphics[width=3.0in]{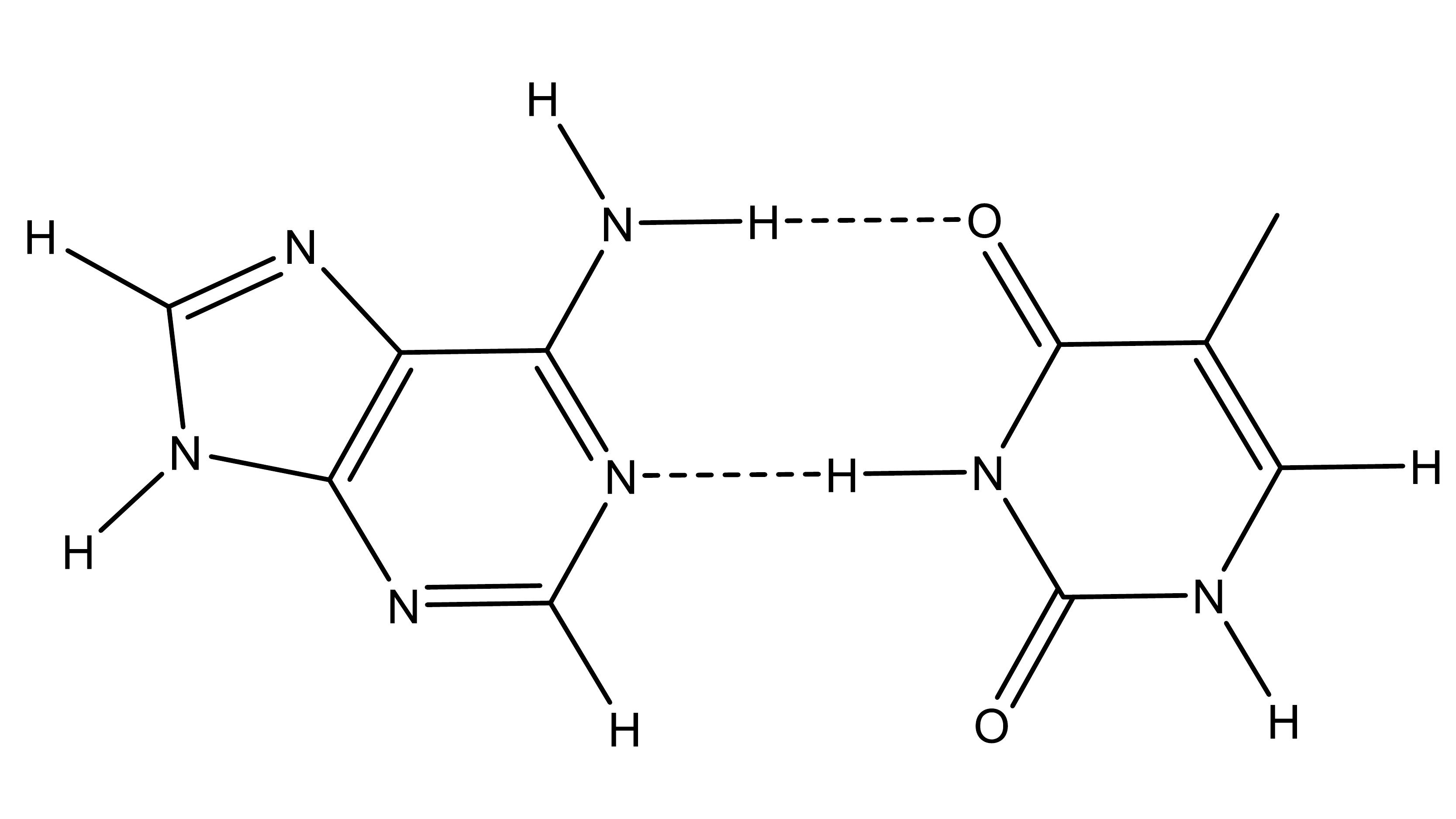}\label{fig:at_structure}}  
	\captionsetup{justification=raggedright,singlelinecheck=false,font=footnotesize}
	\caption{Molecular structures of the guanine--cytosine (a) and adenine--thymine (b) base pairs.
	}
	\label{fig:gc_at_structure}
\end{figure*}

Finally, we apply our efficient density-fitted EA/IP-ADC implementation to compute VEA and AEA of two DNA nucleotide base pairs: guanine--cytosine (GC) and adenine--thymine (AT), shown in \cref{fig:gc_at_structure}.  Investigation of electron attachment to nucleobases is important for understanding light-induced biochemical processes, such as electron or hole transfer along the DNA/RNA strands that can eventually lead to genetic mutation.\cite{Teoule:1987p573,Barrios:2002p7991,Boudaiffa:2000p1658} In our previous work,\cite{Banerjee:2019p224112} we reported VEA's for the five isolated nucleobases [adenine (A), cytosine (C), guanine (G), thymine (T), and uracil (U)] computed using the EA-ADC($n$) ($n$ = 2, 3), CCSD(T), and EOM-CCSD methods. The calculated VEA's of the nucleobases decreased in the order G $>$ U $\gtrsim$ C $\gtrsim$ T $>$ A, in a good agreement with the G $>$ U $\gtrsim$ T $\gtrsim$ C $>$ A trend from the experiment.\cite{Aflatooni:1998p6205} However, the nucleobases primarily appear in pairs, linked via a network of hydrogen bonds that can alter the electron affinities of the nucleobases, making it difficult to predict the site of electron localization in the base pairs.\cite{Colson:1992p9787,Richardson:2002p10163} 

Although no experimental electron affinities have been reported for the isolated base pairs, several theoretical studies have been performed.\cite{Colson:1992p9787,Richardson:2002p10163,Richardson:2003p848,Reynisson:2002p5353,Al:2000p2994,Smets:2001p342,Peng:2012p120,Tripathi:2019p10131} Early computational investigations using Hartree--Fock and density functional theories (DFT)\cite{Colson:1992p9787,Richardson:2002p10163,Richardson:2003p848} reported that the electron attachment to the GC base pair preferentially localizes the electron on the C nucleobase, violating the G $>$ C trend observed for the isolated nucleobases. For the AT base pair, the electron was found to localize on T, in agreement with the T $>$ A trend in the experimental EA's. Another DFT study reported A to be the primary site for the electron localization in the AT anion.\cite{Reynisson:2002p5353} Electron attachment to GC and AT was studied by the Adamowicz\cite{Al:2000p2994,Smets:2001p342} and Schaefer\cite{Peng:2012p120} groups using the MP2 method with the double-zeta basis sets incorporating additional diffuse functions. They reported AEA's of the GC/AT pairs to be $-$0.06/$-$0.4 eV and 0.01/$-$0.42 eV, respectively. Both of these studies found the evidence of electron localization on A in the AT anion. Very recently, Tripathi et al.\@ performed a study of electron attachment to the GC and AT base pairs using the EOM-CCSD method.\cite{Tripathi:2019p10131} Using the aug-cc-pVTZ basis set with additional 5s5p4d diffuse functions, they reported the computed VEA's for the GC/AT pair to be 0.099/0.003 eV and the AEA's of 0.237/$-$0.014 eV, respectively. The authors of this study also demonstrated that at the equilibrium geometry of the neutral base pairs the electron attachment occurs in the dipole-bound state, while it occurs in the valence-bound state at the equilibrium geometries of the AT and GC anions.

\begin{figure*}[t!]
    \subfigure[]{\includegraphics[width=3.0in]{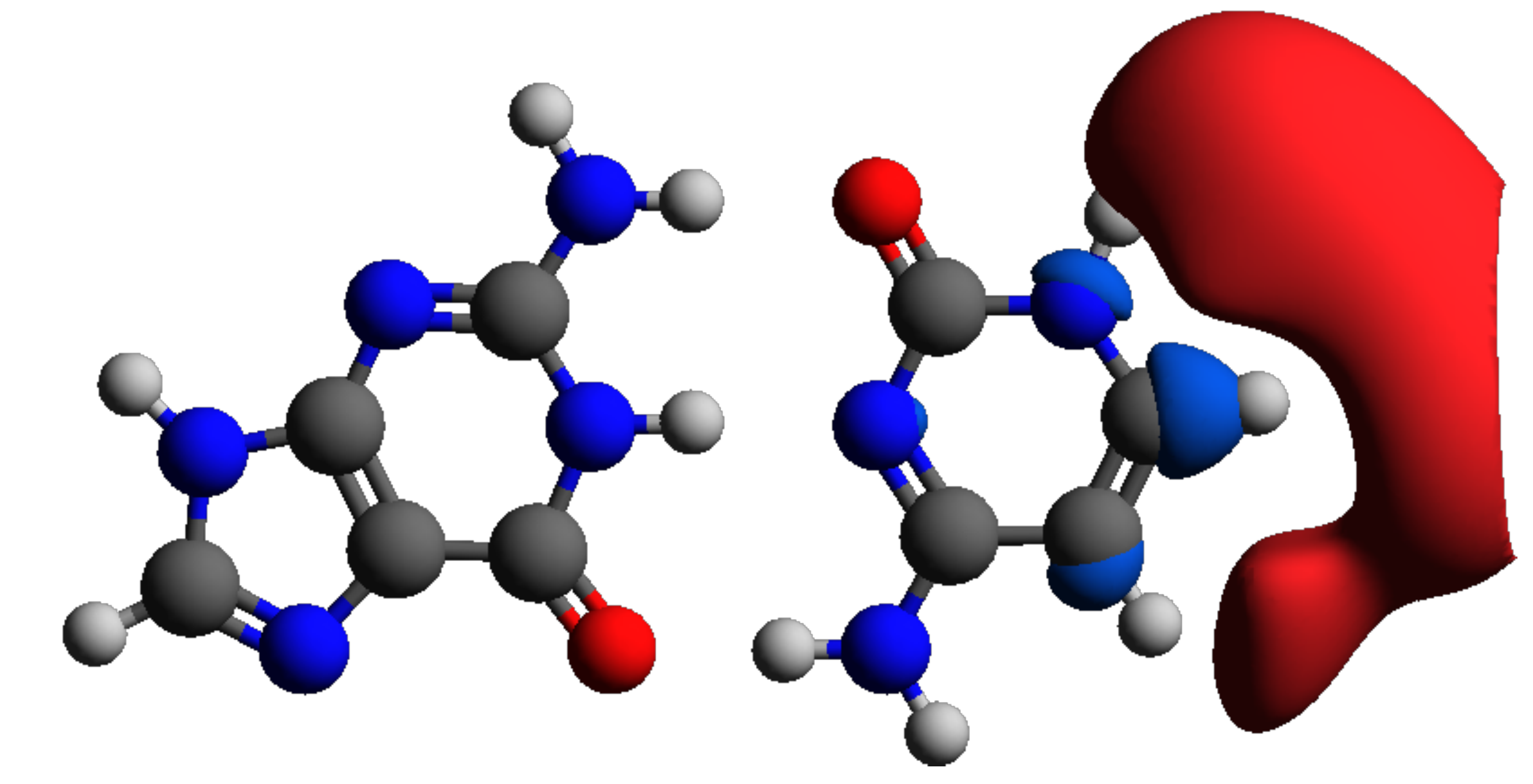}\label{fig:gc_dyson_neutral}} \qquad
    \subfigure[]{\includegraphics[width=3.0in]{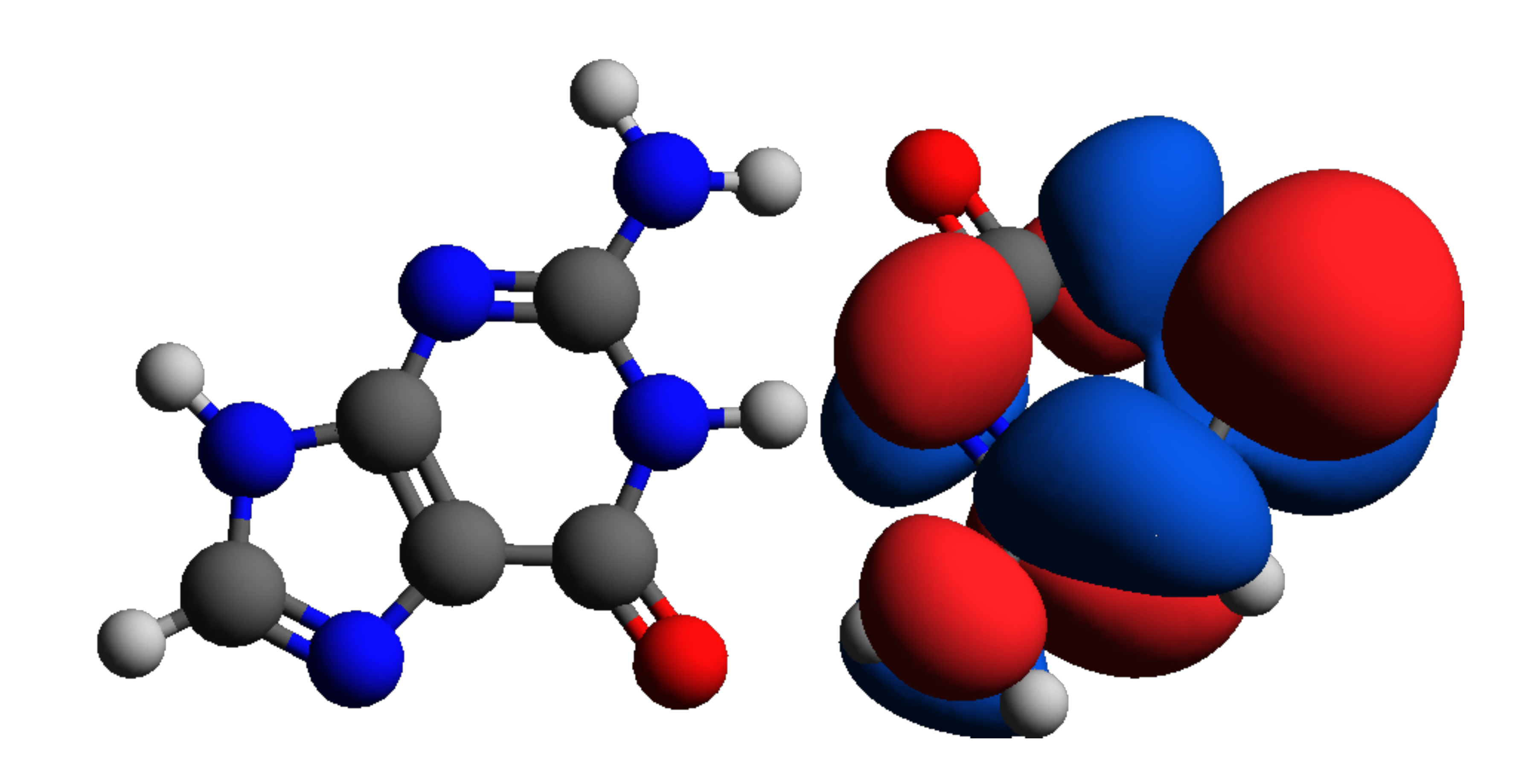}\label{fig:gc_dyson_anion}} 
	\captionsetup{justification=raggedright,singlelinecheck=false,font=footnotesize}
	\caption{Dyson orbitals for the lowest-energy electron attachment to the guanine--cytosine base pair at the neutral (a) and anion (b) equilibrium geometries computed using the EA-ADC(3) method with the aug-cc-pVDZ basis set. 
	}
	\label{fig:gc_dyson}
\end{figure*}

\begin{figure*}[t!]
    \subfigure[]{\includegraphics[width=3.0in]{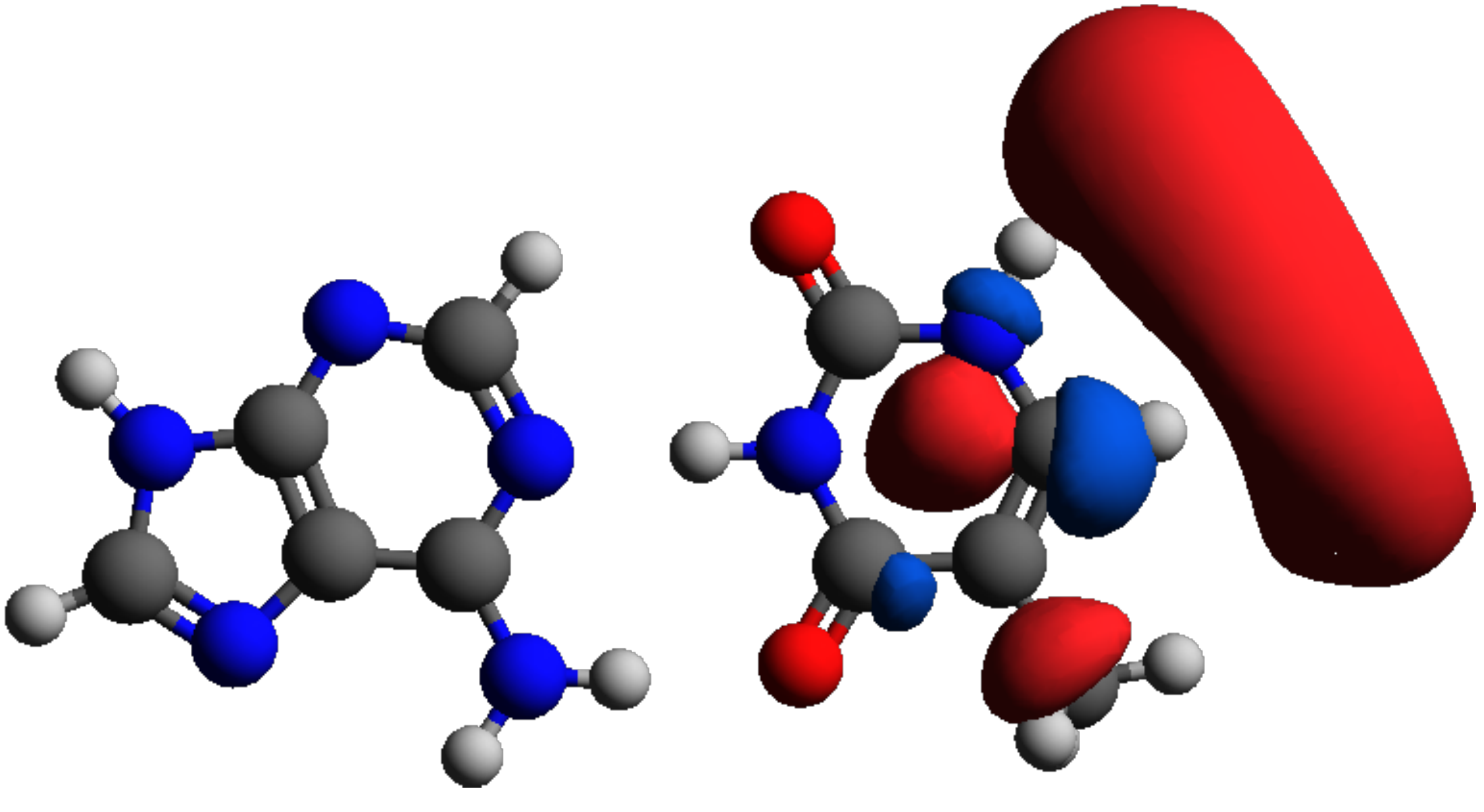}\label{fig:at_dyson_neutral}} \qquad
    \subfigure[]{\includegraphics[width=3.0in]{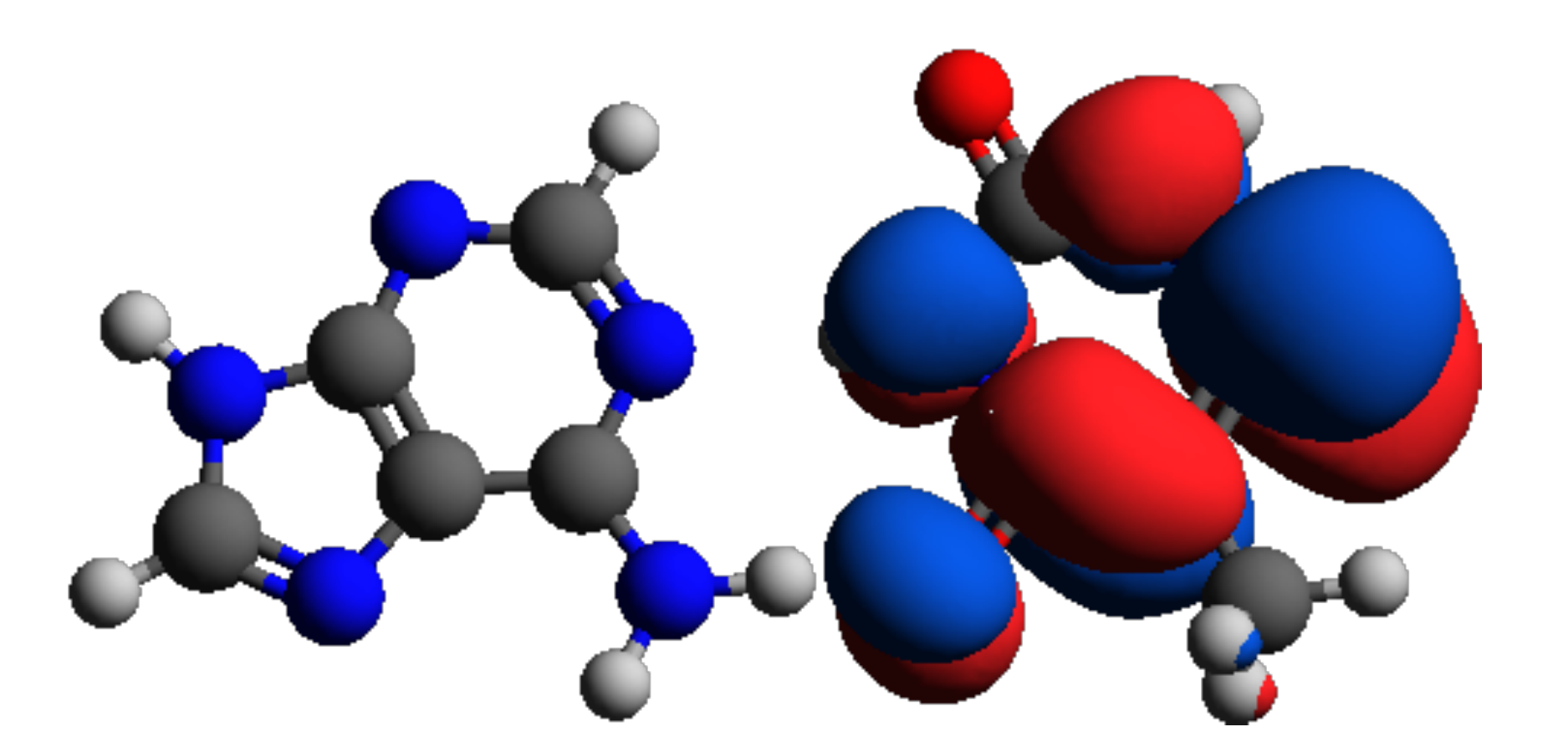}\label{fig:at_dyson_anion}} 
	\captionsetup{justification=raggedright,singlelinecheck=false,font=footnotesize}
	\caption{Dyson orbitals for the lowest-energy electron attachment to the adenine--thymine base pair at the neutral (a) and anion (b) equilibrium geometries computed using the EA-ADC(3) method with the aug-cc-pVDZ basis set. 
	}
	\label{fig:at_dyson}	
\end{figure*}

\cref{fig:gc_dyson,fig:at_dyson} show the EA-ADC(3) Dyson orbitals for the lowest-energy electron attachment to the GC and AT base pairs computed at the neutral (a) and anion (b) equilibrium geometries. In agreement with the results of Tripathi et al.,\cite{Tripathi:2019p10131} the Dyson orbitals computed at the neutral equilibrium geometries are localized outside of each molecule, showing the formation of the dipole-bound states. Relaxation to the anion equilibrium geometries leads to a formation of the valence-bound states with the Dyson orbitals localized on C of GC (\cref{fig:gc_dyson_anion}) and T of AT (\cref{fig:at_dyson_anion}), which agrees well with the results of other post-Hartree--Fock calculations reported in the literature.\cite{Al:2000p2994,Smets:2001p342,Peng:2012p120,Tripathi:2019p10131}

\begin{table*}[t!]
\begin{threeparttable}
	\captionsetup{justification=raggedright,singlelinecheck=false,font=footnotesize}
	\caption{Vertical (VEA) and adiabatic (AEA) electron attachment energies (in eV) of the guanine--cytosine and adenine--thymine base pairs computed using MP$n$ and EA-ADC($n$) ($n$ = 2, 3) methods. Also shown are the EOM-CCSD/aug-cc-pV$X$Z+5s5p4d ($X$ = D, T) reference results from Ref.\@ \citenum{Tripathi:2019p10131}.
	}
	\label{tab:base_pairs}
	\footnotesize
	\setstretch{1}
    \begin{tabular}{L{5.5cm}C{2.6cm}C{2.6cm}C{2.6cm}C{2.6cm}}
       \hline
        \hline
        \multicolumn{1}{c}{Method} &\multicolumn{2}{c}{Guanine--Cytosine} &\multicolumn{2}{c}{Adenine--Thymine} \\
        &VEA  &AEA  &VEA  &AEA  \\
               \hline 
MP2/aug-cc-pVTZ                				&$-$0.43    	&0.07       				        &$-$0.69   		                 &$-$0.29       \\   
MP3/aug(nH)-cc-pVTZ                			&$-$0.40    	&0.09     				        	                                                                                 \\
EA-ADC(2)/aug-cc-pVTZ      				&0.19         	&0.69     	                                 &$-$0.06    	                          &0.34\\
EA-ADC(3)/aug(nH)-cc-pVTZ      			&$-$0.34    	&0.15       	                                 &$-$0.54  		                         &$-$0.14\tnote{a}\\
MP2/d-aug-cc-pVDZ                				&0.06 		&0.05     		                         &$-$0.07                                    &$-$0.33       \\   
MP3/d-aug-cc-pVDZ                				&0.06    		&0.10                                         &$-$0.07	                                 &$-$0.41 \\
EA-ADC(2)/d-aug-cc-pVDZ				&0.14                &0.14        	                        &0.02    			                  &$-$0.24\\
EA-ADC(3)/d-aug-cc-pVDZ				&0.09                &0.13                                        &$-$0.03    	                       &$-$0.38\\     
\hline
EOM-CCSD/aug-cc-pVDZ+5s5p4d\tnote{b}	&0.10      		&0.18 				        &0.00 			               &$-$0.10        \\
EOM-CCSD/aug-cc-pVTZ+5s5p4d\tnote{b}	&0.10      		&0.24 				        &0.00 			               &$-$0.01        \\
\hline
\hline
    \end{tabular}
        \begin{tablenotes}
    \item[a] The AEA was obtained using \cref{eq:aea_from_vea} where $\Delta E_{MPn}$ was computed using MP2/aug-cc-pVTZ.
    \item[b] Ref.\@ \citenum{Tripathi:2019p10131}.
     \end{tablenotes}
\end{threeparttable}
\end{table*}

\cref{tab:base_pairs} presents the vertical (VEA) and adiabatic (AEA) electron affinities of GC and AT computed using the second- and third-order MP$n$ and ADC($n$) methods. Due to the dipole-bound nature of the vertically-attached electronic states, the computed VEA's are expected to exhibit a strong basis set dependence. For this reason, in our calculations we employ two types of basis sets: singly-augmented aug-cc-pVTZ and aug(nH)-cc-pVTZ and doubly-augmented d-aug-cc-pVDZ. Due to the limitations of the available unrestricted MP3 implementations, we do not report VEA and AEA of AT computed at the MP3/aug(nH)-cc-pVTZ level of theory. We compare our results to VEA's and AEA's calculated using the EOM-CCSD method with the aug-cc-pV$X$Z+5s5p4d ($X$ = D, T) basis sets from Ref.\@ \citenum{Tripathi:2019p10131}.

As expected, the computed VEA's show a strong dependence on the basis set. For both GC and AT base pairs, the MP$n$ and ADC($n$) VEA's computed using the d-aug-cc-pVDZ basis set are in a very good agreement with the reference results, while the results obtained using the singly-augmented triple-zeta basis sets show much larger deviations, indicating that the additional diffuse orbitals are critical for accurate predictions of VEA's of these molecules. The best results are demonstrated by the EA-ADC(3)/d-aug-cc-pVDZ method, which predicts VEA's of 0.09/$-$0.03 eV for the GC/AT pair, in a close agreement with the reference VEA's of 0.10/0.00 eV. The EA-ADC(2) method combined with the d-aug-cc-pVDZ basis set shows somewhat larger deviations from the reference results with VEA's of 0.14/0.02 eV. The MP$n$/d-aug-cc-pVDZ ($n$ = 2, 3) methods exhibit larger errors with VEA's of $\sim$ 0.06/$-$0.07 eV.

The computed AEA's exhibit much weaker basis set dependence. Except for EA-ADC(2)/aug-cc-pVTZ, all levels of theory with either of the two basis sets predict a positive AEA for GC and a negative AEA for AT, in agreement with the reference results. For GC, the EA-ADC(3) method shows the best agreement with the reference AEA (0.18 to 0.24 eV) predicting AEA's of 0.15 and 0.13 eV computed using the aug(nH)-cc-pVTZ and d-aug-cc-pVDZ basis sets, respectively. When combined with the aug(nH)-cc-pVTZ basis set, the EA-ADC(3) method yields an accurate AEA of AT ($-$0.14 eV), which agrees well with the reference AEA's in the $-0.10$ to $-0.01$ range. At the EA-ADC(3)/d-aug-cc-pVDZ level of theory, the computed AEA has a lower value of $-$0.38 eV. When using the d-aug-cc-pVDZ basis set, the MP$n$ methods predict AEA's that are close to those of EA-ADC($n$), but further away from the reference. The fact that the VEA's and AEA's computed using EA-ADC($n$) and MP$n$ with the d-aug-cc-pVDZ basis set agree well with the reference results for GC, but show larger deviations for AT suggests that the AEA of the latter base pair requires a higher-level description of electron correlation effects, beyond those included in the finite-order perturbation theories. 

\section{Conclusions}
\label{sec:conclusions}

In this work, we presented an efficient implementation of the single-reference algebraic diagrammatic construction (ADC) theory for simulating electron attachment (EA) and ionization (IP) energies and spectra of molecules (EA/IP-ADC($n$), $n=2,3$). Our new implementation, available in the \textsc{PySCF} program, consists of the spin-restricted (RADC) and unrestricted (UADC) codes for closed- and open-shell systems, respectively. The RADC and UADC programs are capable of efficient memory and disk management by employing the density fitting approximation for the two-electron integrals and take advantage of the highly optimized subroutines for tensor contractions and parallelization. Combining the RADC implementation with density fitting results in a four-fold speedup of the EA/IP-ADC(3) calculations for closed-shell molecules, relative to the UADC implementation with unapproximated two-electron integrals. Our numerical tests show that the density fitting approximation has a negligible effect on the accuracy of the ADC methods but dramatically lowers the computational cost of the integral transformation and the input/output operation count.

We demonstrated the capabilities of our efficient implementation by applying the EA/IP-ADC($n$) ($n=2,3$) methods to the TEMPO radical and the EA-ADC($n$) methods to the guanine--cytosine (GC) and adenine--thymine (AT) DNA base pairs. Our EA- and IP-ADC(3) computations of closed-shell molecules with up to 1028 basis functions and open-shell molecules with up to 758 molecular orbitals are, to the best of our knowledge, the largest calculations using these methods reported to date. For the TEMPO radical, the photoelectron spectra simulated using IP-ADC($n$) ($n=2,3$) are in a good agreement with an experimental spectrum. The EA-ADC($n$) ($n=2,3$) methods predict that the TEMPO radical has a negative vertical electron affinity (VEA), suggesting that the electron attachment is energetically unfavorable at the equilibrium geometry of the neutral radical. Upon structural relaxation, the electron attachment becomes favorable, as indicated by the small positive values (0.08 to 0.49 eV) of the adiabatic electron affinities (AEA) computed using EA-ADC($n$) ($n=2,3$). For the DNA base pairs, our EA-ADC results confirm the dipole-bound nature of the vertically-attached electronic states reported in the earlier EOM-CCSD study.\cite{Tripathi:2019p10131} The VEA's computed using the EA-ADC methods show a strong dependence on the number of diffuse basis functions employed in the calculation. Using the doubly-augmented d-aug-cc-pVDZ basis set, the EA-ADC($n$) ($n=2,3$) VEA's are in a good agreement with the available reference data. The computed AEA's show a weaker basis set dependence. For GC, both EA-ADC(2) and EA-ADC(3) predict accurate AEA's ($\sim$ 0.14 eV), which agree well with AEA from EOM-CCSD (0.18 to 0.24 eV). The computed AEA of AT show larger errors ($\sim$ 0.15 to 0.30 eV), highlighting the importance of infinite-order electron correlation effects for the electron affinity of this system.

In summary, our work demonstrates that EA/IP-ADC($n$) ($n=2,3$) are efficient and accurate theoretical methods for simulations of charged excitations and spectra that can be routinely applied to closed- and open-shell molecules with more than 1000 basis functions. Our efficient implementation enables new applications that were previously out of reach and can be combined with periodic boundary conditions for simulations of crystalline chemical systems. Work along this direction is ongoing in our group.

\section{Supplementary Material}
See supplementary material for the optimized Cartesian geometries of the TEMPO radical, DNA base pairs, tables showing the density fitting approximation errors for EA/IP-DF-ADC($n$) ($n$ = 2, 3), and tabulated photoelectron spectra of the TEMPO radical computed using IP-UADC($n$) ($n$ = 2, 3).

\section{Acknowledgements}
This work was supported by the start-up funds provided by the Ohio State University. Additionally, S.B.\@ was supported by a fellowship from the Molecular Sciences Software Institute under NSF Grant ACI-1547580. Computations were performed at the Ohio Supercomputer Center under projects PAS1583.\cite{OhioSupercomputerCenter1987} 

\section{Data availability}
The data that supports the findings of this study are available within the article and its supplementary material. Additional data can be made available upon reasonable request.

\section{Appendix: Calculation of spectroscopic factors in the RADC implementation}
\label{sec:Appendix}
Here, we demonstrate how to calculate the spectroscopic factors in the RADC implementation. As discussed in \cref{sec:implementation:spin-adaptation}, solution of the RADC eigenvalue problem \eqref{eq: eig} delivers the spin-adapted eigenvectors $\mathbf{Y_\pm}$. Since the RADC eigenvectors contain fewer elements than those in the UADC implementation, they need to be renormalized to ensure that the computed spectroscopic factors are consistent with those calculated using UADC. For IP-RADC, the elements of the renormalized eigenvectors $\mathbf{\tilde{Y}}_-$ can be obtained from $\mathbf{Y_-}$ as follows:
\begin{align}
\tilde{Y}_{-i,\mu} &= \frac{Y_{-i,\mu}}{\sqrt {Y^{norm}_{-\mu}}} \label{eq:appendix_1} \\
\tilde{Y}_{-ija,\mu} &= \frac{Y_{-ija,\mu}}{\sqrt {Y^{norm}_{-\mu}}} \label{eq:appendix_2}
\end{align}
where $Y_{-\mu}^{norm}$ is a norm of root $\mu$, defined as
\begin{align}
Y_{-\mu}^{norm} &= \sum_{i} Y_{-i,\mu}Y_{-i,\mu} \notag \\
&+ \sum_{ija} (2 Y_{-ija,\mu} Y_{-ija,\mu} - Y_{-ija,\mu} Y_{-jia,\mu}) \label{eq:appendix_3}
\end{align}
Once computed, the renormalized eigenvectors $\mathbf{\tilde{Y}}_{-}$ are contracted with the elements of the spin-adapted transition moments matrix $\mathbf{{T}}_{-}$ to obtain the spin-adapted spectroscopic amplitudes ($\mathbf{{X}}_{-}$)
\begin{align}
X_{-p\mu} &= \sum_{i} T_{-p,i}\tilde{Y}_{-i,\mu} \notag \\ 
&+ \sum_{ija} (2T_{-p,ija}\tilde{Y}_{-ija,\mu}  - T_{-p,jia}\tilde{Y}_{-ija,\mu}) \label{eq:appendix_4}
\end{align}
For EA-RADC, expressions for the elements of the renormalized eigenvectors $\mathbf{\tilde{Y}}_+$ and spectroscopic amplitudes $\mathbf{{X}}_{+}$ can be obtained from \cref{eq:appendix_1,eq:appendix_2,eq:appendix_3,eq:appendix_4} by replacing the 1h ($i$) and 2h-1p ($ija$) excitation labels with those of the 1p ($a$) and 2p-1h ($abi$) excitations. Using the elements of $\mathbf{{X}}_\pm$, the spectroscopic factors can be obtained as follows
\begin{align}
P_{\pm \mu} &= 2 \sum_{p} |X_{\pm p\mu}|^2
\end{align}

\end{document}